\documentclass[useAMS,usenatbib]{mn2e}
\usepackage{graphicx, subfig, amssymb, amsmath,rotating}

\def\kpnothanks{Visiting Astronomer, Kitt Peak National Observatory, National Optical Astronomy Observatory,
which is operated by the Association of Universities for Research in Astronomy (AURA) under cooperative
agreement with the National Science Foundation.}

\title[Kinematics of M31 outer halo GCs]
  {The outer halo globular cluster system of M31 -- II. Kinematics}
\author[Veljanoski et al.] {J.~Veljanoski$^1$, A. D.~Mackey$^2$, A. M. N.~Ferguson$^{1,}$\thanks{\kpnothanks}, A. P.~Huxor$^3$, P.~C\^{o}t\'{e}$^{4,\star}$\,
\newauthor M. J.~Irwin$^5$, N. R.~Tanvir$^6$, J.~Pe\~{n}arrubia$^1$, E. J.~Bernard$^1$, M.~Fardal$^7$, N. F.~Martin$^{8,9}$,
\newauthor A.~McConnachie$^4$, G. F.~Lewis$^{10}$, S. C.~Chapman$^{11}$, R. A.~Ibata$^8$, A.~Babul$^{12}$
\\
$^1$ Institute for Astronomy, University of Edinburgh, Royal Observatory, Blackford Hill, Edinburgh, EH9 3HJ, UK\\
$^2$ Research School of Astronomy $\&$ Astrophysics, Australian National University, Mt. Stromlo Observatory, Cotter Road,\\ Weston Creek, ACT 2611, Australia\\
$^3$ Astronomisches Rechen-Institut, Zentrum f\"{u}r Astronomie der Universit\"{a}t Heidelberg, M\"{o}nchhofstra{\ss}e 12-14, 69120 Heidelberg, \\Germany.\\
$^4$ NRC Herzberg Institute of Astrophysics, 5071 West Saanich Road, Victoria, BC, V9E 2E7, Canada \\
$^5$ Institute of Astronomy, University of Cambridge, Madingley Road, Cambridge, CB3 0HA, UK \\
$^6$ Department of Physics $\&$ Astronomy, University of Leicester, Leicester LE1 7RH, UK \\
$^7$ University of Massachusetts, Department of Astronomy, LGRT 619-E, 710 N. Pleasant Street, Amherst, Massachusetts, 01003-9305, USA \\
$^8$ Observatoire de Strasbourg, 11, rue de l'Universit\'e, F-67000 Strasbourg, France \\
$^9$ Max-Planck-Institut fuer Astronomie, Koenigstuhl 17, D-69117 Heidelberg, Germany \\
$^10$ Institute of Astronomy, School of Physics, University of Sydney, NSW 2006, Australia \\
$^11$ Dalhousie University, Dept. of Physics and Atmospheric Science, Coburg Road Halifax, B3H1A6, Canada \\
$^12$ Department of Physics and Astronomy, University of Victoria, Elliott Building, 3800 Finnerty Road Victoria, BC V8P 5C2 Canada }

\pagerange{\pageref{firstpage}--\pageref{lastpage}} \pubyear{2014}

\def\LaTeX{L\kern-.36em\raise.3ex\hbox{a}\kern-.15em
    T\kern-.1667em\lower.7ex\hbox{E}\kern-.125emX}

\begin{document}

\label{firstpage}

\maketitle

\begin{abstract}
We present a detailed kinematic analysis of the outer halo globular cluster system of the Andromeda galaxy (M31).
Our basis for this is a set of new spectroscopic observations for 78 clusters lying at projected distances between
$R_{\rm proj} \sim20$-$140$ kpc from the M31 centre. These are largely drawn from the recent PAndAS globular cluster
catalogue; 63 of our targets have no previous velocity data. Via a Bayesian maximum likelihood analysis
we find that globular clusters with $R_{\rm proj} > 30$ kpc exhibit coherent rotation around the minor optical axis of M31,
in the same direction as more centrally-located globular clusters, but with a smaller amplitude of
$86 \pm 17$ km$\;$s$^{-1}$. There is also evidence that the velocity dispersion of the outer halo globular
cluster system decreases as a function of projected distance from the M31 centre, and that this relation can be
well described by a power law of index $\approx -0.5$. The velocity dispersion profile of the outer
halo globular clusters is quite similar to that of the halo stars, at least out to the radius up to which there is available
information on the stellar kinematics.
We detect and discuss various velocity correlations amongst subgroups of globular clusters that lie on stellar debris
streams in the M31 halo. Many of these subgroups are dynamically cold, exhibiting internal velocity dispersions consistent
with zero. Simple Monte Carlo experiments imply that such configurations are unlikely to form by chance,
adding weight to the notion that a significant fraction of the outer halo globular clusters in M31 have been accreted alongside their
parent dwarf galaxies. We also estimate the M31 mass within 200 kpc via the Tracer Mass Estimator (TME), finding
$(1.2 - 1.6)\pm0.2 \times 10^{12}M_{\odot}$. This quantity is subject to additional
systematic effects due to various limitations of the data, and assumptions built in into the TME.
Finally, we discuss our results in the context of formation scenarios for the M31 halo.
\end{abstract}

\begin{keywords}
Local Group --- galaxies: individual (M31) --- galaxies: kinematics and dynamics --- galaxies: halos --- globular clusters: general
\end{keywords}

\section{Introduction}
Even after decades of active research, our understanding of the various processes governing
galaxy formation and evolution remains incomplete. In the currently favoured $\Lambda$CDM cosmological models and
their semi-analytic extensions, the extended dark matter and stellar haloes of galaxies are at least partly
formed through hierarchical build-up of smaller fragments, akin to the dwarf galaxies we observe today
\citep[e.g.][]{Abadi03a,Abadi03b,Purcell07,QiGuo11}. Observations of various stellar streams
in the halo of our Galaxy \citep[e.g.][]{Belokurov06,Grillmair06a,Grillmair06b,Martin14}, many of which are thought
to be tidally disrupted dwarf galaxies, support this idea. Indeed, observations of the Sagittarius dwarf
galaxy, which is currently being accreted onto the Milky Way, provide direct evidence that this process
is still on-going \citep[e.g.][]{Ibata94,Majewski03,Koposov12,Slater13}.
However, searching for and studying such features is challenging. Halo streams are typically
very faint and therefore difficult to observe even in the Milky Way. Seeking the progenitor systems
of tidal streams is also complicated, because satellite galaxies that are losing stellar mass to
tides tend to become faint and cold \citep{Penarrubia08}, and because their survival time depends
strongly on both their (typically unknown) mass distribution \citep{Penarrubia10} and the shape
of the dark matter halo of their host \citep{Penarrubia02}.

Another way to probe galaxy haloes is through their globular cluster (GC) systems. Due to
their high luminosities, GCs are observed much more easily than the underlying stellar field
components in these remote parts of galaxies. Various studies have found correlations between GC systems
and their host galaxy properties that can shed light on galaxy formation mechanisms \citep{GCReview06}.
Since GCs are frequently found at large radii, their kinematics constitute a particularly useful tool.
GC motions provide information about the assembly history of the host galaxy, its
total mass, the shape of the gravitational potential and the dark matter distribution
\citep[e.g.][]{Schuberth10,Schuberth12,Strader11}.  Of relevance to the present paper is that
GCs also provide an alternative way to look for and study past accretion events, by searching for spatially
and dynamically linked GC groups that can serve as tracer populations for their (now disrupted)
parent systems.

In the Milky Way, various properties of the halo
GCs such as their ages, horizontal branch morphologies, luminosities, sizes and kinematics, are
consistent with them having an external origin, in line with expectations from hierarchical formation
models \citep[e.g.][]{SZ78,Mackey04,Mackey05,Franch09,Forbes10b,Keller12}. Indeed, it has been
clearly demonstrated that a number of outer halo GCs are kinematically associated with the Sagittarius stream
\citep[e.g.,][]{Ibata95,DaCosta95}. However, despite this, and despite the fact that GCs provided the
first clues that the Milky Way halo formed at least to some degree through the coalescence of smaller
fragments \citep{SZ78}, it has proven difficult to establish the existence of dynamically linked Galactic GC groups
\citep[e.g.,][]{Palma02}, or locate additional examples where Galactic GCs are clearly kinematically
(or even spatially) associated with stellar streams from the disruption of their host dwarfs.

The Andromeda galaxy (M31) provides a unique opportunity for detailed study of galaxy assembly
processes. Its close proximity of $\sim780$ kpc \citep{Conn12} makes M31 the only massive spiral
galaxy other than our own in which both star clusters and the diffuse stellar field can be resolved into
individual stars. Subtending a large area on the sky, M31 provides a much clearer view of a typical
spiral galaxy halo than our own Galaxy, where one must observe vast angular regions and battle
with projection effects and hugely variable extinction. M31 hosts a rich GC system, with over 450
confirmed members listed in the Revised Bologna Catalogue\footnote{http://www.bo.astro.it/M31/}
\citep{GalletiRBC04}, most of which lie within 30 kpc in projection from the galactic centre.

Kinematics have been of particular interest for studies of the M31 GC system. The first
radial velocities for M31 GCs were obtained by \citet{vdBergh69}. Later \citet{Hartwick74} used the
available data to estimate the mass of M31 using its GCs as kinematic tracers. A number of works
followed \citep{Huchra82,Federici90,Huchra91,Federici93}, increasing the GC velocity data set,
updating the kinematic mass estimate and providing velocity dispersions for different subsamples.
Using higher precision data for over 200 GCs, all having projected radii smaller than 22 kpc,
\citet{Perrett02} showed that unlike in the Milky Way, the GC system of M31 exhibits a strong
rotation around the minor optical axis of the galaxy. More recent investigations have further
enlarged the total number of radial velocity measurements, and presented updated kinematic
analyses and mass estimates \citep{Galleti06,Galleti07,Lee08,Caldwell11}. It is important to emphasise
that all these past investigations have focused on GCs at projected distances less than 30 kpc from the M31 centre.

A number of studies have used the available kinematic data to search for and detect possible velocity
sub-clustering amongst the M31 GCs, which in turn may reflect past mergers or accretions in the system.
The first significant attempt was by \citet{Ashman93}, who used a technique in which groups of GCs were
isolated based on deviations of the global mean velocity and velocity dispersion between each cluster and its
$N$ nearest companions. This technique yielded a number of groups, but \citet{Ashman93} warned that their
method may produce false positives if the GC system were to exhibit significant rotation, which was later found to
indeed be the case \citep{Perrett02}. Using an improved and enlarged data set, \citet{Perrett03} searched for
sub-clustering in the inner M31 GC system as evidence of past merger remnants. These authors employed a
modified friends-of-friends algorithm which can detect the elongated groups that are expected to be found
along tidal debris streams. \citet{Perrett03} detected 10 unique groups of at least 4 GCs in each. They
performed additional tests and found that even though the majority of these might be chance groupings, there
was a high probability that at least some might be genuine dynamically linked units. More recently,
\citet{Perina09} attempted to identify clusters in the inner parts of M31 sitting away from the global trend
in metallicity with position. They located three such GCs, sitting at similar projected radii and possessing
matching velocities quite distinct from the kinematics of the M31 disk, altogether suggestive of being physically
part of a coherent structure.

We again emphasise that these previous attempts at finding coherent GC groups in M31 were undertaken for
objects lying at projected radii $R_{\rm proj}\lesssim30$ kpc, where such searches are extremely challenging.
It is difficult, if not impossible, to trace a single stellar debris stream reliably due to the presence of many
intertwined stellar substructures, as well as the comparatively high stellar densities of the M31 spheroid.
In addition, the high number density and wide range of GC properties makes it difficult for distinct kinematic
groups to be robustly detected. Because the dynamical timescales in the inner parts of M31 are also
comparatively short, it is also likely that any accreted objects presently found in these regions are now
well mixed with the host population of stars and clusters and no longer retain their initial kinematic
relationships.

In recent years, our international collaboration has initiated a sequence of wide-area imaging surveys
in order to explore in detail the far outer halo of M31, culminating in the recent Pan-Andromeda Archaeological
Survey \citep[PAndAS;][]{McConnachie09}. Our high quality data has enabled us to search for, and on account of their
partially resolved nature, unambiguously detect GCs out to a projected radius ($R_{\rm proj}$) of $\approx140$ kpc
\citep{Huxor05,Huxor08,Huxor14}, and, in at least one case, a 3D radius of $\approx 200$ kpc
\citep{Martin06,Mackey10a}. In total, our group has discovered 80 GCs in M31 with galactocentric distances
larger than 30 kpc in projection. For comparison, prior to our work only 3 globular clusters were known at
such large radii. One particularly interesting conclusion drawn from the results of the PAndAS survey is that
a large fraction ($\sim\,$50-80\%) of these remote GCs preferentially lie projected on top of various stellar streams
and other tidal debris features uncovered in the M31 halo \citep{Mackey10b}. Monte Carlo simulations suggest
that the probability of such alignment arising by chance is lower than 1\%. This finding is a major step forward
in understanding how the M31 outer halo GC population formed, and supports the idea first put forward by
\citet{Cote00a} that accretion was the dominant mode of formation for this GC system.

In a recent Letter \citep{Veljanoski13a}, we presented an initial survey of the kinematics of
the M31 outer halo GC system. We showed for the first time that the outer halo GC system of
M31 appears to be rotating and that there is a hint of a decreasing velocity dispersion amongst the
population as a function of increasing projected radius.

In the present contribution, we significantly increase the GC sample for which radial velocities have
been measured, and present the first detailed kinematic analysis of the outer halo GC system of M31.
This paper is structured in the following manner. Section 2 contains a complete description of the data, the data
reduction, and our methodology for determining radial velocities. In Section 3 we use a Bayesian framework
to derive the global kinematic properties of the M31 outer halo GCs, while in Section 4 we focus on
various velocity correlations and sub-clustering observed for GCs that lie along particular stellar debris features.
Finally, we discuss the implications of our results in Section 5, followed by the summary in Section 6.

\section{Observations and data reduction}
\label{s:obs}

\subsection{The sample}
Our spectroscopy campaign involves eight separate observing runs
conducted with three different facilities: the ISIS spectrograph mounted on the
4.2m William Herschel Telescope (WHT), the RC spectrograph on the 4.0m Mayall
Telescope at the Kitt Peak National Observatory (KPNO), and the GMOS instrument installed on the
8.1m Gemini-North telescope. The targets were selected from a catalogue of outer halo M31
GCs comprised mainly of objects discovered in the PAndAS survey \citep[][Paper I in this series]{Huxor14},
but also including clusters found in previous searches by our group
\citep{Martin06,Huxor08}, as well as objects listed in the Revised
Bologna Catalogue \citep[RBC,][]{GalletiRBC04}. In total our list of possible targets
contained 83 GCs situated at $R_{\rm proj}$ larger than 30 kpc, plus those at smaller radii
presented in Paper I. Table \ref{tab:obslog} shows the log of observations for all eight observing
runs. Altogether we made 90 separate GC observations, acquiring spectra for 78 different clusters,
of which 63 had no previous spectroscopic information. Repeated observations of some GCs
were made primarily to facilitate consistency checks but also to supplement lower
quality data in a few instances.

\begin{figure}
\begin{center}
\end{center}
\includegraphics[width=86mm]{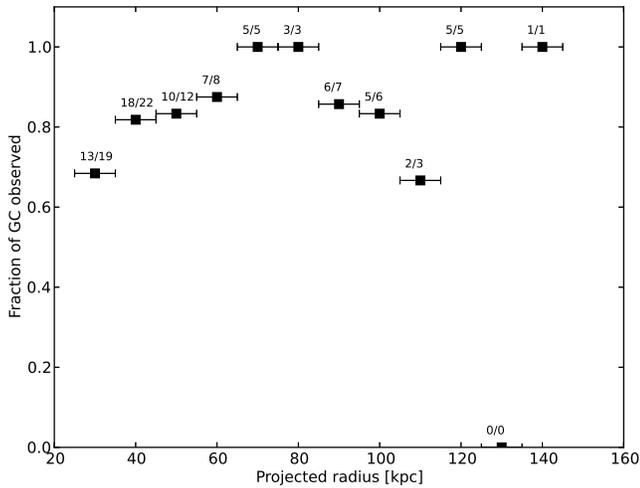}
\caption{The fraction of GCs in our measured sample as a function of
$R_{\rm proj}$, between 25 and 145 kpc in bins of 10 kpc.}
\label{fig:compl}
\end{figure}

Throughout the campaign, priority was given to clusters lying on top of stellar substructures,
and to those having larger $R_{\rm proj}$. Our final observed sample consists of GCs with $R_{\rm proj}$
between 18 and 141 kpc.  Most, however, lie beyond 30 kpc in projection from the centre of M31 -- a region
which, throughout the remainder of this paper, we will refer to as the ``outer halo". This radius
corresponds to the clear break in the GC radial number density profile observed by
\citet[][see also Mackey et al. 2014, in prep.]{Huxor11}. In total we acquired spectra for 71 clusters
with $R_{\rm proj} > 30$ kpc, corresponding to 85.5\% of the known globular clusters in the M31 outer halo.
Of these, there are 20 in our sample beyond 80 kpc including 10 beyond 100 kpc. The full radial
distribution of our observed clusters is shown in Figure \ref{fig:compl}.

In the following subsections we describe the data and the data
reduction process. Even though the reduction procedure is standard and
similar in the case of the 4m and 8m class telescopes, we discuss it
separately in order to highlight any differences.

\begin{table*}
 \caption{Observing log showing the instruments used, the dates of observation,
	  the program numbers, the principal investigator of each observation proposal,
          the observation modes and the number of GCs observed in each run. Note
          that certain globular clusters were repeatedly observed in different observing runs.}
 \label{tab:obslog}
 \begin{tabular}{llllll}
 	\hline
	\hline
	Instrument		& Date of obs.		& Program Number	& PI  	& Obs. mode 	& No. GCs	\\
	\hline
	WHT/ISIS		& 29/09-02/10 2005	& .			& A. P. Huxor		& Visitor	& 19 		\\
	WHT/ISIS 		& 16/08-18/08 2009	& .			& A. P. Huxor		& Visitor	& 12 		\\
	WHT/ISIS		& 09/09-11/09 2010	& .			& A. P. Huxor		& Visitor	& 13 		\\
	KPNO/RC			& 13/08-17/08 2009	& .			& A. M. N. Ferguson	& Visitor	& 17 		\\
	Gemini/GMOS-N		& 20/07-02/09 2010	& GN-2010B-Q-19		& A. D. Mackey		& Service 	& 4  		\\
	Gemini/GMOS-N		& 02/08-05/09 2011	& GN-2011B-Q-61		& A. D. Mackey		& Service 	& 11 		\\
	Gemini/GMOS-N		& 29/07-13/09 2012	& GN-2012B-Q-77		& A. D. Mackey		& Service 	& 7  		\\
	Gemini/GMOS-N		& 02/08-31/08 2013	& GN-2013B-Q-66		& A. D. Mackey		& Service 	& 7  		\\
	\hline
 \end{tabular}
\end{table*}

\subsection{WHT and KPNO data}
We used the ISIS instrument mounted on the 4.2m WHT for three observing runs,
performing longslit spectroscopic observations of 41 different GCs in our sample. ISIS has
two detectors (``arms''), that independently sample two separate wavelength ranges,
a bluer and a redder one. In all runs we set the slit width at 1.5 - 2$^{\prime\prime}$. For the
blue arm we used the R600B grating to cover the wavelength range between $\sim$3500 and 5100 \AA,
and the EEV12 detector with a dispersion of 0.45 \AA\ pixel$^{-1}$.  The spectral resolution was
R $\sim 1500$. For the red arm we used the R600B grating covering the wavelength range
between $\sim$7500 - 9200 \AA, and the REDPLUS camera, achieving a dispersion of 0.49
\AA\ pixel$^{-1}$. The spectral resolution was R $\sim 2700$. The only exception to this
set-up was for the observing run conducted in 2005, when only the blue arm of ISIS was used.
We observed each GC as a series of short exposures, with the total integration
time varying between 600 and 7200 seconds depending on the cluster brightness.
The data are unbinned in both the spatial and wavelength direction. The typical signal-to-noise
of the spectra is $\sim 7 - 20$ per \AA, while reaching $\sim 70$ per \AA\ for the brightest targets.

We used the RC spectrograph mounted on the KPNO-4m telescope in single
slit mode to obtain spectra of 17 GCs during a single observing run in 2009.
For this purpose, we used a slit width of 2$''$ and the T2KB detector, along with the KPC007 grating
to select the wavelength range $\sim$3500 - 6500 \AA\ with a dispersion of 0.139
\AA\ pixel$^{-1}$ and a spectral resolution of R $\sim 1300$. A similar observing strategy
to that which we used for the WHT observations was adopted. Each cluster was
observed with multiple exposures, and the total integration time ranged
between 600 and 6400 seconds depending on the brightness of the target.
There is no binning of the data in either the wavelength
or the spatial directions. The typical signal to noise of the spectra is 25 - 50 per \AA.

The data obtained with the WHT and the KPNO telescopes were reduced using standard {\sc iraf}\footnote{
  {\sc iraf} is distributed  by the National Optical Astronomy Observatories, which are
  operated  by the Association of Universities for Research in Astronomy, Inc., under
  cooperative agreement with the National Science Foundation.}
procedures. The basic reduction of the spectra (bias and overscan subtraction, flat-fielding,
illumination correction) was done with dedicated standard tasks, which are part of the \textsc{ccdred}
package. The \emph{apall} task in the \textsc{kpnoslit} package was used to extract one
dimensional spectra from the two-dimensional frames. The extraction apertures had radii
of 2 - 2.5$''$. The same task was also used interactively to select background sky regions
and to find the trace. The sky in the selected regions of the target spectra was then fit
with a 2$^{\rm nd}$ order Chebyshev polynomial and subtracted. The spectra were traced using a 3$^{\rm rd}$ order
cubic spline function, and were extracted using the optimal variance weighting option in \emph{apall}.
An advantage of \emph{apall} is that it also produces an error spectrum based on
the Poisson noise of the spectrum that is being extracted.

Wavelength calibration was based on Cu-Ne-Ar and He-Ne-Ar lamps for WHT and KPNO
spectra respectively. Comparison ``arcs'' were obtained before and after each program
target exposure. The arc spectra were extracted using the same \emph{apall} parameters
as the target GCs they were used to calibrate. The \emph{identify} task was used to identify
$\sim50$ RC, $\sim90$ ISIS blue and $\sim25$ ISIS red lines in the arc spectra, and the
dispersion solution was fit with a 3$^{\rm rd}$ order cubic-spline function. The RMS residuals of
the fits were 0.08 $\pm$ 0.01 \AA\, 0.05 $\pm$ 0.01 \AA\ and 0.02 $\pm$ 0.01 \AA\
for the data obtained with the RC, ISIS blue arm and ISIS red arm instruments respectively.
Since two wavelength solutions were found for each target from the `before' and `after' arcs,
they were averaged and assigned to the GC spectrum via the \emph{dispcor} task. To check whether
the wavelength calibration is reliable, we measured the positions of sky lines in separately
extracted sky spectra. We found that the wavelength calibration is accurate to 0.08 \AA\
with no systematic shifts for all data observed with the 4m class telescopes.

\begin{figure*}
\begin{center}
\end{center}
\includegraphics[width=151mm, angle = 0]{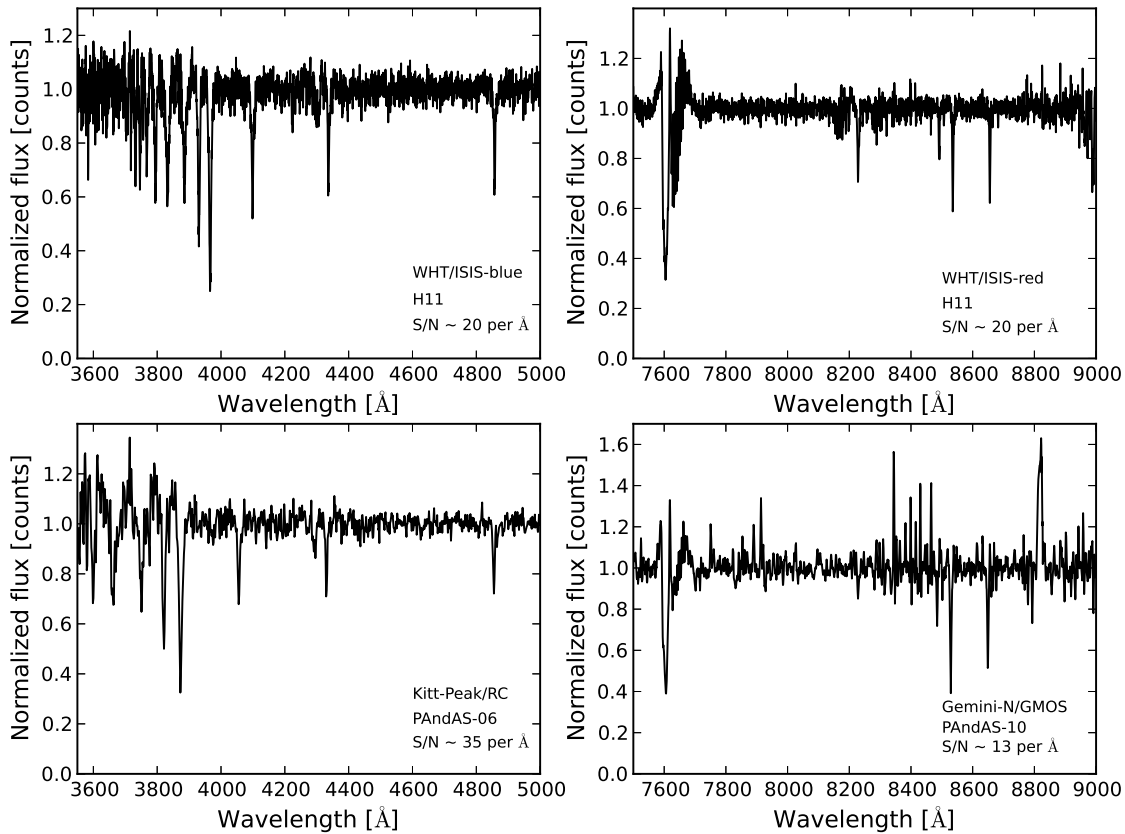}
\caption{Typical continuum normalised spectra obtained with each of the employed
instruments in our spectroscopy campaign. Note that the sky subtraction residuals
in the GMOS spectra are larger in a relative sense due to the faintness of those targets.}
\label{fig:specExamples}
\end{figure*}

The multiple exposures of each target GC from a given observing run were stacked in the
following manner. First, all exposures were shifted into the heliocentric frame and interpolated onto
a common wavelength scale.
They were then combined together as described by Equation \ref{eq:stack}:
\begin{equation}
{\rm S}_{i} = \frac{\sum_{j} \frac{{\rm S}_{i,j}}{\eta^{2}_{i,j}}}{\sum_{j}\frac{1}{\eta^{2}_{i,j}}}
\label{eq:stack}
\end{equation}
where S represents a spectrum, while $\eta$ is the corresponding error spectrum.
The index $i$ corresponds to a particular pixel in an exposure $j$. Finally the
spectra were continuum subtracted, for the purpose of measuring radial velocities.
Several examples of fully reduced spectra are shown in Figure \ref{fig:specExamples}.
The displayed spectra are continuum normalized rather than continuum subtracted in order to preserve the
relative strengths of the absorption features for better visualization.

\subsection{Gemini data}
To observe the fainter as well as the more diffuse and extended GCs, which
typically have lower surface brightness than classical compact globular clusters, we used the
GMOS instrument mounted on the Gemini-North telescope. Spectra were taken for a total of
29 objects over the course of four separate observing runs executed in service mode between
2010 and 2013. The observations were conducted using a longslit mask with a slit width of 0.75$''$.
The grating of choice was R831, which was used to cover the wavelength range between
$\sim$ 7450 and 9500 \AA. To account for the gaps between the chips of the GMOS detector,
two or three sets of three exposures were taken with slightly different grating angles and hence
slightly different central wavelengths. For each science exposure in each
set, we dithered the telescope by $\pm$15$''$ along the spatial direction of the slit.
This helps to minimise any effects coming from imperfections in the detector as well as
any systematic effects of the background sky subtraction. The typical total integration
time was $\sim$5700 seconds. The data were binned by a factor of two in the wavelength direction,
obtaining a resolution of 0.68 \AA\ pixel$^{-1}$, as well as in the spatial
direction to achieve a spatial resolution of $\sim$0.147$''$ per pixel. The spectral resolution
achieved with this setup was R $\sim 4000$. The average signal to noise of the data is
$\sim 15$ per \AA.

The data reduction was undertaken using {\sc iraf}, employing tasks from the dedicated
\textsc{gemini/gmos} package provided and maintained by the Gemini staff. The reduction was
carried out separately for data sets observed with different central wavelengths. A master
bias frame was created with the \emph{gbias} task from $>30$ raw bias frames acquired near to
the time of the program observations. The standard overscan and bias subtraction, flat-fielding
and mosaicing of the three chips of the detector into a single frame was done with the
\emph{gsreduce} task. Unlike the data taken with the 4m class telescopes, here we
wavelength calibrated the two dimensional frames before one-dimensional spectra were
extracted. The wavelength calibration is based on Cu-Ar arcs, and such frames
were taken before and after each set of program target exposures, with the central
wavelength of the arcs matching the central wavelength of the observed GCs in a
specific set. The \emph{wavelength} task was employed to identify $\sim16$ lines
in the arc spectrum and to fit a dispersion solution using a fourth order Chebyshev
polynomial. The typical RMS of the fit is 0.02 \AA. The \emph{gstransform} task was
employed to assign a wavelength solution to each GC frame.

\begin{table*}
 \caption{Information regarding the radial velocity standard stars and globular clusters used. (1)
	Star/Cluster ID, (2) Right Ascension, (3) Declination, (4) Spectral type of the stars, (5)
	Heliocentric radial velocity, (6) number of exposures, (7) Instrument used, (8) Year of
	observation and (9) Source of the heliocentric radial velocity. }
 \label{tab:rvstd}
 \begin{tabular}{lccclclcl}
 	\hline
	\hline
	ID		& \multicolumn{2}{c}{Position (J2000.0)}	& Spec.	& $V_{\rm helio}$ 		& No. 	& Instrument	& Year 	& Reference		\\
			& 	RA		& 	Dec		& Type	& [km$\;$s$^{-1}$]	& Exp.	&			&		&			\\
	\hline
	HD 4388		& 00 46 27.0	& +30 57 05.6	& K3III		& -27.5 $\pm$ 0.3	& 4		& KPNO/RC	& 2009	& \citet{Udry99}	\\
			&		&		&		&			& 1		& WHT/ISIS	& 2009	&			\\
			&		&		&		&			& 3		& WHT/ISIS	& 2010	&			\\
	\hline
	HD 12029	& 01 58 41.9 	& +29 22 47.7	& K2III		& 38.5 $\pm$ 0.3	& 4		& KPNO/RC	& 2009	& \citet{Udry99}	\\
	\hline
	HD 145001	& 16 08 04.5	& +17 02 49.1	& G8III		& -10.3 $\pm$ 0.3	& 4		& KPNO/RC	& 2009	& \citet{Udry99}	\\
	\hline
	HD 149803	& 16 35 54.3	& +29 44 43.3	& F7V		& -7.5 $\pm$ 0.7	& 1		& KPNO/RC	& 2009	& \citet{Udry99}	\\
	\hline
	HD 154417	& 17 05 16.8	& +00 42 09.2	& F9V		& -18.6 $\pm$ 0.3	& 7		& KPNO/RC	& 2009	& \citet{Udry99}	\\
			&		&		&		&			& 5		& WHT/ISIS	& 2009	&			\\
			&		&		&		&			& 5		& WHT/ISIS	& 2010	&			\\
	\hline
	HD 171391	& 18 35 02.4	& -10 58 37.9	& G8III		& 7.4 $\pm$ 0.2		& 5		& KPNO/RC	& 2009	& \citet{Udry99}	\\
	\hline
	G1		& 00 32 46.8 	& +39 34 42.6 	& 		& -332 $\pm$ 3		& 1 		& KPNO/RC	& 2009	& \citet{GalletiRBC04}	\\
			&		&		&		&			& 2		& WHT/ISIS	& 2005	&			\\
			&		&		&		&			& 4		& WHT/ISIS	& 2009	&			\\
			&		&		&		&			& 3		& WHT/ISIS	& 2010	&			\\
			&		&		&		&			& 9		& Gemini/GMOS-N	& 2010	&			\\
	\hline
	MGC1		& 00 50 42.4 	& +32 54 58.7 	& 		& -355 $\pm$ 2		& 2		& WHT/ISIS	& 2009	& \citet{Alves-Brito09}	\\
			&		&		&		&			& 3		& WHT/ISIS	& 2010	&			\\
			&		&		&		&			& 9		& Gemini/GMOS-N	& 2011	&			\\
	\hline
 \end{tabular}
\end{table*}

One dimensional spectra were extracted with the standard {\sc iraf} \emph{apall} package.
As the Gemini observations primarily targeted faint and diffuse clusters, often
multiple apertures were used to extract the light coming from individual bright
stars within the cluster. Typical effective aperture radii range from 0.7 to 2.5$''$.
The internal velocity dispersion of extended and low luminosity clusters is smaller
than the measured radial velocity uncertainty of each star in such a cluster, which
makes this approach an appropriate one.
The \emph{apall} task was also used to subtract the background sky, and to find the
trace, in a similar way as for the 4m data. The extraction was done with the
variance weighting option on. Finally, all exposures for a given GC observation
were stacked together in the same manner as for the 4m telescope data.
If multiple extraction apertures were used, they were also stacked together following
the prescription in Equation \ref{eq:stack}, producing a final one dimensional science
spectrum. These science spectra were then continuum subtracted as required
for measuring radial velocities. An example of a representative, fully reduced spectrum is
shown in Figure \ref{fig:specExamples}.

\subsection{Radial velocity measurements}
For the purpose of determining the radial velocities of the GCs, throughout our
observing campaign we also performed multiple observations of 6 different radial
velocity standard stars. The standard stars were chosen to have a stable and accurately
known radial velocity, to be of a certain spectral type so their spectra would be similar to the
GC spectra, and to be sufficiently bright so a high S/N spectrum could be obtained
with a very short exposure.

In addition we also used two M31 GCs, G1 and MGC1, as radial velocity
templates. These clusters have high precision radial velocity information obtained from high
resolution spectra \citep{GalletiRBC04,Alves-Brito09}, comprise some of the brightest GCs in
the outskirts of M31, and possess metallicities spanning the expected range for outer halo GCs
$-2.3 \la [$Fe$/$H$] \la -1.0$. These properties allow for high S/N spectra to be observed with
relatively short exposures and makes the two GCs appropriate radial velocity templates.
Table \ref{tab:rvstd} displays the relevant data regarding the radial velocity standard stars and
template GCs.

Heliocentric radial velocities were determined via a customized routine that performs
a $\chi^2$ minimisation between the target and template spectra.
First, the template is adjusted to the wavelength scale of the target spectrum.
The template spectrum is then Doppler shifted by an input velocity, which is systematically
varied between -1000 to 500 km s$^{-1}$ in increments of 1 km s$^{-1}$ in the heliocentric
frame. The chosen velocity search range is large enough to comfortably encompass the
expected velocities of all GCs that belong to the M31 system. The $\chi^2$ match
between the target and template spectra is then calculated via:
\begin{equation}
\chi^{2} = \sum_{i}\frac{({\rm {d}}_{i}-{k{\rm M}({\vartriangle}v, \sigma)}_{i})^{2}}{\eta_{i}^{2}+\delta_{i}^{2}}
\label{eq:chi2}
\end{equation}
where $i$ is the pixel index, d is the spectrum of the target GC (`data' in the statistical sense),
and M is the template spectrum (the `model' against which the data are tested).
The uncertainties in the target and template spectra are $\eta$ and $\delta$
respectively. The model M is a function of two free parameters. The first is the input
velocity ${\vartriangle}v$. The second parameter, $\sigma$, is due to the different width
of the absorption lines in the target cluster and the template star spectrum, caused by the
internal velocity dispersion of the stars that comprise a certain GC. However, as the resolution
of the spectrographs we have employed is not sufficient to probe the internal velocity
dispersions of the GCs, this parameter can be ignored. The parameter $k$ accounts for the flux
difference between the target and the template spectra. It is not independent, and can be calculated
via:
\begin{equation}
k = \frac{\sum_{i}\frac{\rm d_{i}M_{i}}{\eta_{i}^{2}+\delta_{i}^{2}}}
{\sum_{i}\frac{\rm{M}_{i}^{2}}{\eta_{i}^{2}+\delta_{i}^{2}}}
\label{eq:k}
\end{equation}
where the symbols are as in Equation \ref{eq:chi2}.
The input velocity corresponding to the minimum of the $\chi^2$ function is the
measured velocity of the GC. In order to remove the large telluric features, any regions of higher
sky subtraction residuals, and the edges of the spectra where the S/N is low, we selected
certain wavelength ranges over which the $\chi^2$ function was calculated. For the data
observed with GMOS-N and the red arm of ISIS, the $\chi^2$ window was selected just around the
Ca$\,${\sc ii} triplet (CaT) lines with range of  8400 - 8750 \AA. For the KPNO data the $\chi^2$
window was in the range of  3831 - 6000 \AA, and for the data observed with the blue arm of ISIS
the corresponding $\chi^2$ window was 3900 - 4900 \AA.

This technique yielded smaller velocity uncertainties by 23\% on average compared to the
results coming from the standard cross correlation method. Usage of the uncertainties in both
the template and target spectra helps to eliminate spurious peaks in the $\chi^2$
function that might arise due to imperfectly subtracted sky lines, which are especially strong
for the faint GCs near the CaT.

\begin{figure}
\begin{center}
\end{center}
\includegraphics[width=86mm]{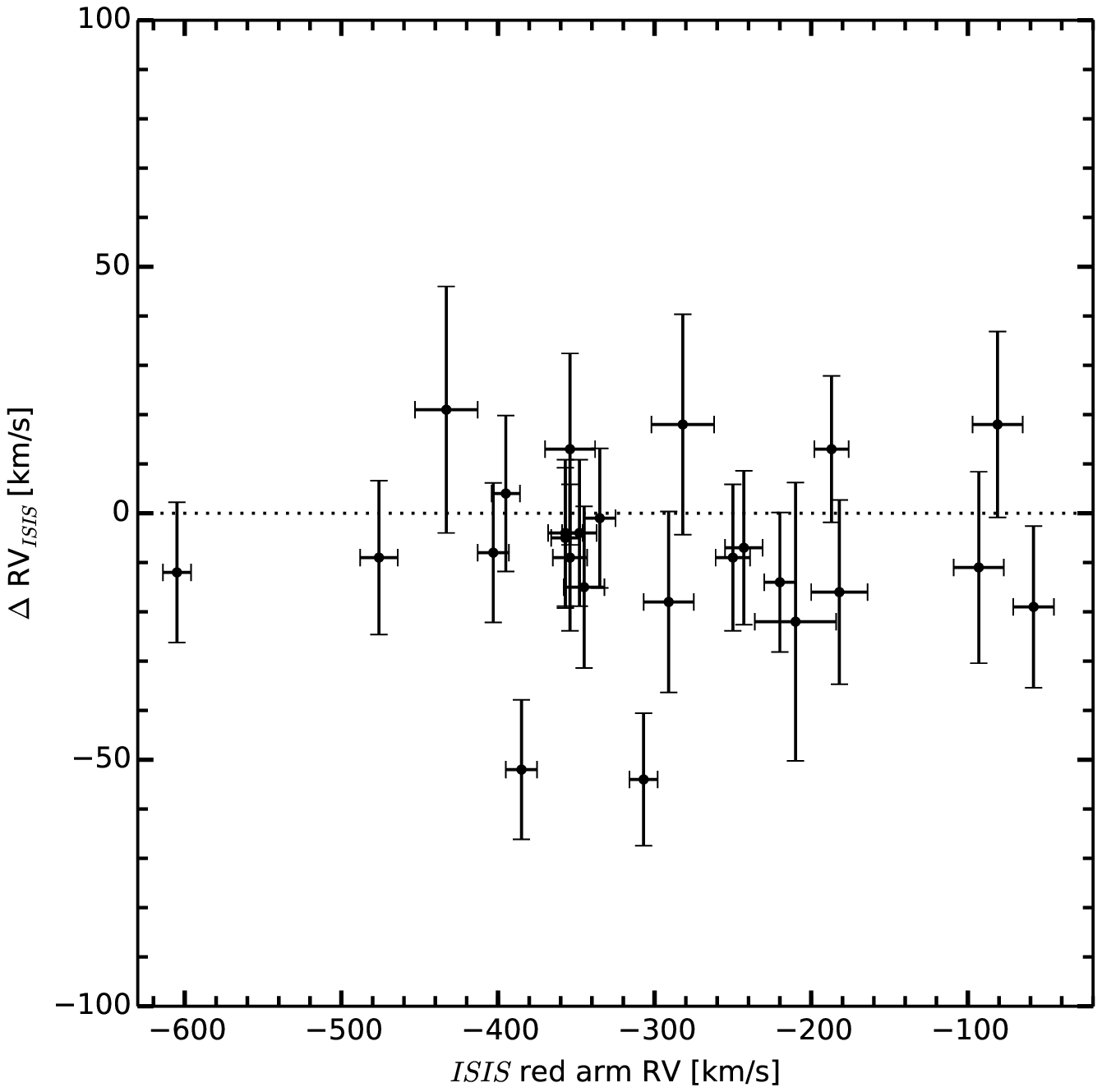}
\caption{Comparison between the radial velocities determined via
  the red and blue arms of ISIS. The dotted black line represents the ideal zero offset.
  It is seen that there is good agreement without any systematic offsets between the two
  independent sets of measurements.}
\label{fig:ISIScomp}
\end{figure}

\begin{table*}
 \caption{Literature radial velocities for the 15 GCs in our sample that were
 previously observed as part of other studies.}
 \label{tab:rvcomp}
 \begin{tabular}{lccccl}
 	\hline
	\hline
        ID                & Alternative ID     & \multicolumn{2}{c}{Position (J2000.0)}        & $V_{\rm helio}$                & Reference \\
                          &                    & RA          &  Dec                       & [km$\;$s$^{-1}$]                &              \\
        \hline
        G1                & Mayall-II          & 00 32 46.5  & +39 34 40 & -332 $\pm$ 3                    & average, see \citet{GalletiRBC04}     \\
        G2                & Mayall-III         & 00 33 33.7  & +39 31 18 & -313 $\pm$ 17                   & average, see \citet{Galleti06}        \\
        B514              & MCGC4              & 00 31 09.8  & +37 54 00 & -458 $\pm$ 23                   & \citet{Galleti07}     \\
        B517              & .                  & 00 59 59.9  & +41 54 06 & -272 $\pm$ 54                   & \citet{Galleti09}     \\
        G339              & BA30               & 00 47 50.2  & +43 09 16 & 33 $\pm$ 30                     & \citet{Federici93}    \\
        EXT8              & .                  & 00 53 14.5  & +41 33 24 & -152 $\pm$ 30                   & \citet{Federici93}    \\
        H1                & MCGC1/B520         & 00 26 47.7  & +39 44 46 & -219 $\pm$ 15                   & \citet{Galleti07}     \\
        H10               & MCGC5              & 00 35 59.7  & +35 41 03 & -358 $\pm$ 2                    & \citet{Alves-Brito09} \\
        H14               & MCGC7              & 00 38 49.4  & +42 22 47 & -248 $\pm$ 24                   & \citet{Caldwell11}    \\
        H23               & MCGC8              & 00 54 24.9  & +39 42 55 & -381 $\pm$ 15                   & Galleti et al, in preparation \\
        H24               & MCGC9              & 00 55 43.9  & +42 46 15 & -147 $\pm$ 20                   & Galleti et al, in preparation \\
        H27               & MCGC10             & 01 07 26.3  & +35 46 48 & -291 $\pm$ 2                    & \citet{Alves-Brito09} \\
        MGC1              & .                  & 00 50 42.4  & +32 54 58 & -355 $\pm$ 2                    & \citet{Alves-Brito09} \\
        PAndAS-07         & PA-7               & 00 10 51.3  & +39 35 58 & -433 $\pm$ 8                    & \citet{Mackey13}      \\
        PAndAS-08         & PA-8               & 00 12 52.4  & +38 17 47 & -411 $\pm$ 4                    & \citet{Mackey13}      \\
	\hline
 \end{tabular}
\end{table*}

\begin{figure}
\begin{center}
\end{center}
\includegraphics[width=86mm]{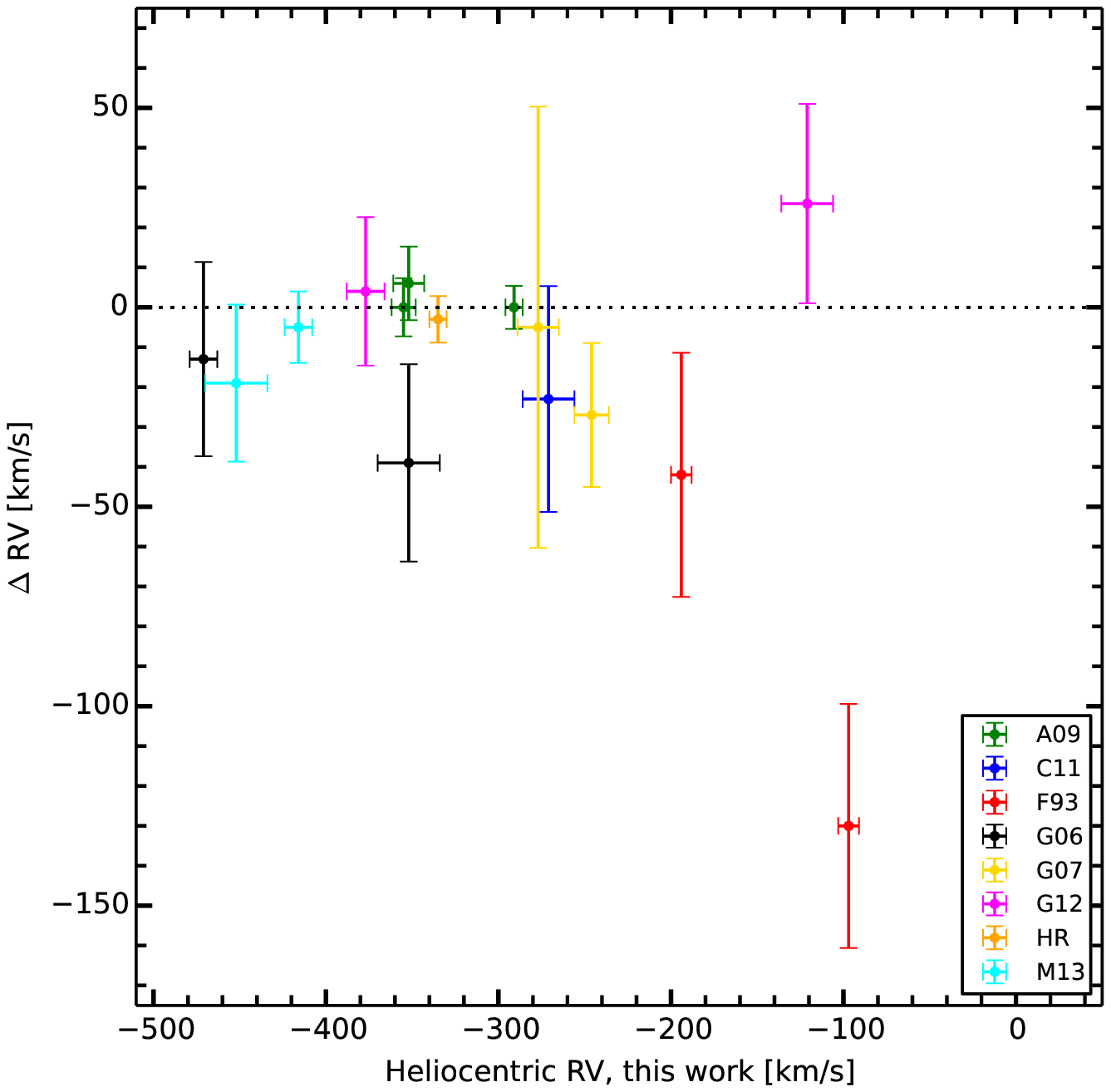}
\caption{Comparison between the heliocentric radial velocities measured in this study to
  those found in the literature for the 15 GCs that have previous observations. The ideal zero
  offset is represented by the black dotted line. Excluding the two most deviant
  points coming from \citet{Federici93}, we find a mean offset of $-8$~km s$^{-1}$, which is
  significantly less than the accompanied RMS scatter of 18~km s$^{-1}$. Hence we see excellent
  agreement between our velocity measurements and those collated in the RBC.
  Adopting the notation from the RBC, the legend key is:
  A09 = \citet{Alves-Brito09};
  C11 = \citet{Caldwell11};
  F93 = \citet{Federici93};
  G06 = \citet{Galleti06};
  G07 = \citet{Galleti07};
  G12 = Galleti et al, in preparation;
  HR = \citet{Peterson89} and \citet{Dubath97};
  M13 = \citet{Mackey13}.}
\label{fig:RVLitComp}
\end{figure}

For GCs observed in a single observing run, the final
adopted radial velocity and its corresponding uncertainty are given by the mean and standard deviation
from all individually obtained velocities resulting from the $\chi^2$ minimisation between
the spectrum of that cluster and all available template spectra, respectively. Regarding the GCs
which were observed with WHT in 2009 and 2010, two independent radial velocities were
measured from the blue and the red arms of ISIS. It is important to note that
these measurements are consistent with each other, and there is no systematic offset between
them. A comparison between these two independent sets of measurements is shown in
Figure \ref{fig:ISIScomp}. {The mean offset is $-8$~km s$^{-1}$, which is much smaller than
the RMS deviation, found to be 20 km s$^{-1}$. Because of the excellent agreement between them, to
obtain final velocities for objects in the 2009 and 2010 WHT runs we simply computed the
error-weighted average of the blue and red arm measurements.

There are 7 clusters that were repeatedly observed in different observing runs. For the radial velocity
of those objects we adopt the error weighted mean of the velocities measured in each run individually,
including the measurements conducted with the two ISIS arms if available.

A total of 15 objects in our sample, listed in Table \ref{tab:rvcomp}, have measured radial
velocities in the literature. We show a comparison between these values and the velocity
measurements from our present study in Figure \ref{fig:RVLitComp}. The mean offset is
$-18$~km s$^{-1}$, while the RMS deviation is found to be 39~km s$^{-1}$. Removing the two
velocity measurements coming from \citet{Federici93}, which is the set of points that are the most
deviant, results in a mean offset of $-8$~km s$^{-1}$ with a RMS scatter of 18~km s$^{-1}$. The
origin of the discrepancy between our velocity measurements and those from \citet{Federici93} is
unclear, but may stem from the very different measurement and calibration techniques employed
in the two different studies. Apart from this, we find excellent agreement between the velocities
derived in this work and those present in the literature, with no signficant systematic offset between
our radial velocity measurements and those collated in the RBC.

\subsection{Corrections for perspective}
\label{ss:persp}
Because our GC sample is spread over a large area of sky spanning $\sim 20\degr$, we converted our
measured radial velocities from the heliocentric to the Galactocentric frame in order to remove any effects
the solar motion could have on the kinematic analysis. The conversion was computed via the relation
found in \citet{Courteau99}, with updated values of the solar motion from \citet{McMillan11} and
\citet{Schonrich10}:
\begin{align}
V_{\rm gal} = V_{\rm helio} + 251.24 \sin(l) \cos(b)  \nonumber \\
+11.1 \cos(l) \cos(b) + 7.25 \sin(b)
\label{eq:gal2hel}
\end{align}
where $l$ and $b$ are the Galactic latitude and longitude.

The wide angular span of our GC sample on the sky introduces additional factors that must be considered.
As per \citet{Vandermarel08}, the observed (Galactocentric) line-of-sight velocity for a target that is part of the
extended M31 system, but separated from its centre by an angle $\rho$ on the sky, can be decomposed as:
\begin{equation}
V_{\rm gal} = V_{\rm M31,r} \cos(\rho) + V_{\rm M31,t} \sin(\rho) \cos(\phi - \theta_{\rm t}) + V_{\rm pec,los}
\label{eq:persp-corr}
\end{equation}
Here, we take M31 to have a systemic radial velocity (measured along the line-of-sight to its centre) of
$V_{\rm M31,r}$, and a systemic transverse velocity $V_{\rm M31,t}$ in a direction on the sky given
by the position angle $\theta_{\rm t}$. The position angle of the target with respect to the centre of M31 is $\phi$,
while $V_{\rm pec,los}$ is its peculiar line-of-sight velocity.

The first two terms in Equation \ref{eq:persp-corr} tell us that with increasing separation $\rho$, a decreasing
fraction of the systemic M31 radial velocity is observed along the line-of-sight to the target, but an increasing fraction
of the transverse motion is carried on this vector. This induces the appearance of a solid body rotation for targets
at wide separations from the centre of the system, around an axis sitting $90\degr$
away from the direction of the transverse velocity on the sky.

One of the main things we wish to test in this paper is whether the outer halo GC system of M31 exhibits
coherent rotation as suggested by \citet{Veljanoski13a}. It is therefore important to consider whether we need
to make a correction for the ``perspective rotation" described above. The most precise measurement of
the M31 transverse velocity to date comes from \citet{Vandermarel12} who found $V_{\rm M31,t} = 17.0$
km s$^{-1}$ with respect to the Milky Way, at a position angle $\theta_{\rm t} \approx 287\degr$ east of north.
Their $1\sigma$ confidence region is $V_{\rm M31,t} \le 34.3$ km s$^{-1}$, consistent with M31 being on a
completely radial orbit towards our Galaxy. The small transverse velocity of M31 means that the induced
perspective rotation for our GC sample is negligible -- at most a few km s$^{-1}$ even for the most remote objects
(which have $\rho \approx 10\degr$). This is smaller than our typical measurement uncertainties.
In principle we could, for completeness, still use Equation \ref{eq:persp-corr} to correct for the rotation; however
the formal uncertainties on the individual components of the \citet{Vandermarel12} transverse velocity
(i.e., the components in the north and west directions on the sky) are $\approx 30$ km s$^{-1}$ each.
Hence making the correction would introduce significantly larger random uncertainties into our final velocity
measurements than ignoring this effect entirely.

We use Equation \ref{eq:persp-corr} with the second term set to zero to obtain the peculiar line-of-sight velocity
of each GC in our sample.  That is, we remove the component due to the radial systemic motion of M31 by
solving for $V_{\rm pec,los}$. In this study we adopt a heliocentric velocity of -301 $\pm$ 1 km s$^{-1}$
for M31 \citep{Vandermarel08}, which corresponds to a Galactocentric radial velocity $V_{\rm M31,r} =$ -109 $\pm$ 4 km s$^{-1}$
\citep[see also][]{Vandermarel12}.

One remaining effect to consider is that each of our final corrected velocities lies along a slightly different
vector due to the different lines-of-sight to the individual GCs in our sample. In principle we ought to correct
these to lie parallel -- adopting the direction of the line-of-sight to the centre of the galaxy might be a logical choice --
before assessing, for example, the how the one-dimensional velocity dispersion of the system varies with projected
radius. However to make this correction for a given GC requires knowledge of its peculiar proper motion, because,
in analogy with Equation \ref{eq:persp-corr}, a small
component of this transverse velocity is carried onto the new vector. In the absence of this information we choose
to leave our measurements unaltered; in any case the expected magnitude of the corrections is, for most targets,
smaller than the random uncertainties on our velocities.

\subsection{Summary}
Table \ref{tab:rvtable} lists the radial velocity measurements for all GCs in our sample. For each object the
raw heliocentric velocity $V_{\rm helio}$ is reported, followed by the Galactocentric velocity $V_{\rm gal}$ from
Equation \ref{eq:gal2hel}, and the peculiar line-of-sight velocity in the M31 frame obtained by solving
Equation \ref{eq:persp-corr} as described towards the end of Section \ref{ss:persp}. We hereafter refer to this
latter velocity as $V_{\rm M31corr}$.

The results of our measurements are also shown in Figure \ref{fig:rvbigmap}, where the observed GCs
from this study are overlaid as coloured points on top of the most recent PAndAS metal-poor
([Fe/H] $\lesssim -1.4$) red giant branch stellar density map. The colour of each marked GC
corresponds to $V_{\rm M31corr}$.

As described in the next Section, we will concentrate our kinematic analysis
on the region outside $R_{\rm proj} = 30$\ kpc. For completeness we searched through the RBC and found that there is
only one cluster with $R_{\rm proj} > 30$ kpc that we have not observed but which has a radial
velocity in the literature. To improve statistics, we add this cluster -- dubbed HEC12, or
alternatively EC4 \citep{Collins09} -- to our sample, and list its relevant data in Table
\ref{tab:rvtable}. Therefore, our final sample of outer halo GCs ($R_{\rm proj} > 30$~kpc)
contains 72 objects in total.

\begin{table*}
\scriptsize
 \caption{Coordinates, projected radius, position angle, angular separation, and heliocentric,
          Galactocentric, M31-corrected and rotation-corrected velocities for the GCs in our sample.
          Clusters for which there exist more accurate radial velocity measurements in the
          literature are marked: (1) from the RBC; (2) from \citet{Mackey13}; ($^\star$) object not
          observed in any of our runs. The instrument abbreviations are W for WHT, K for KPNO and
          G for Gemini-N.}
 \label{tab:rvtable}
 \begin{tabular}{lccccccccccl}
        \hline
        \hline
        Cluster ID              & \multicolumn{2}{c}{Position (J2000.0)} & $R_{\rm proj}$       & PA & $\rho$ & $V_{\rm helio}$  & $V_{\rm gal}$         & $V_{\rm M31corr}$ & $V_{\rm rot-corr}$    & Prominent     & Instrument \\
                                & RA                &  Dec                                  &  [kpc] & [deg] & [deg] & [km s$^{-1}$] & [km s$^{-1}$] & [km s$^{-1}$]  &   [km s$^{-1}$]              & substructure  &             \\
        \hline
        B514 &                  00 31 09.8 &         +37 54 00 &  55.2 &   214.4     & 4.04 &   -471 $\pm$ 8 &      -279 $\pm$ 8 &      -169 $\pm$ 8 &      -84 &          .                     & W            \\
        B517 &                  00 59 59.9 &         +41 54 06 &  44.9 &    77.5     & 3.40 &   -277 $\pm$ 13 &     -93 $\pm$ 13 &      16 $\pm$ 13 &       -57 &          Stream C/D (overlap)  & K            \\
        EXT8 &                  00 53 14.5 &         +41 33 24 &  27.2 &     80.8    & 1.99 &   -194 $\pm$ 6 &      -7 $\pm$ 7 &        102 $\pm$ 7 &           &          .                     & W            \\
        G001$^1$ &              00 32 46.5 &         +39 34 40 &  34.7 &    229.1    & 2.54 &   -335 $\pm$ 5 &      -141 $\pm$ 6 &      -31 $\pm$ 6 &        58 &          Association 2         & W, K, G            \\
        G002$^1$ &              00 33 33.7 &         +39 31 18 &  33.8 &     225.7   & 2.47 &   -352 $\pm$ 19 &     -158 $\pm$ 19 &     -49 $\pm$ 19 &       77 &          Association 2         & G            \\
        G268 &                  00 44 10.0 &         +42 46 57 &  21.0 &    9.8      & 1.54 &   -277 $\pm$ 8 &      -84 $\pm$ 8 &       25 $\pm$ 8 &            &          .                     & W            \\
        G339 &                  00 47 50.2 &         +43 09 16 &  28.8 &     26.2    & 2.11 &   -97 $\pm$ 6 &       95 $\pm$ 6 &        204 $\pm$ 6 &           &          .                     & W            \\
        H1 &                    00 26 47.7 &         +39 44 46 &  46.3 &    244.6    & 3.39 &   -245 $\pm$ 7 &      -48 $\pm$ 7 &       61 $\pm$ 7 &        143 &          .                     & W            \\
        H2 &                    00 28 03.2 &         +40 02 55 &  41.6 &    247.5    & 3.04 &   -519 $\pm$ 16 &     -322 $\pm$ 16 &     -212 $\pm$ 16 &    -133 &          Association 2         & W            \\
        H3 &                    00 29 30.1 &         +41 50 31 &  34.7 &   284.1     & 2.54 &   -86 $\pm$ 9 &       113 $\pm$ 9 &       221 $\pm$ 9 &       266 &          .                     & W            \\
        H4 &                    00 29 44.9 &         +41 13 09 &  33.4 &    269.9    & 2.44 &   -368 $\pm$ 8 &      -170 $\pm$ 8 &      -61 $\pm$ 8 &         1 &          .                     & W            \\
        H5 &                    00 30 27.2 &         +41 36 19 &  31.8 &     279.3   & 2.33 &   -392 $\pm$ 12 &     -194 $\pm$ 12 &     -85 $\pm$ 12 &      -34 &          .                     & W            \\
        H7 &                    00 31 54.5 &         +40 06 47 &  32.2 &    241.5    & 2.24 &   -426 $\pm$ 23 &     -231 $\pm$ 23 &     -121 $\pm$ 23 &     -38 &          Association 2         & K            \\
        H8 &                    00 34 15.4 &         +39 52 53 &  29.1 &   229.9     & 2.13 &   -463 $\pm$ 3 &      -269 $\pm$ 4 &      -160 $\pm$ 4 &      -73 &          Association 2         & G            \\
        H9 &                    00 34 17.2 &         +37 30 43 &  56.1 &   204.2     & 4.10 &   -374 $\pm$ 5 &      -184 $\pm$ 6 &      -74 $\pm$ 6 &         7 &          .                     & W            \\
        H10$^1$ &               00 35 59.7 &         +35 41 03 &  78.4 &   193.8     & 5.47 &   -352 $\pm$ 9 &      -165 $\pm$ 9 &      -56 $\pm$ 9 &        12 &          .                     & W            \\
        H11 &                   00 37 28.0 &         +44 11 26 &  42.1 &   342.1     & 3.08 &   -213 $\pm$ 7 &      -15 $\pm$ 7 &       93 $\pm$ 7 &         54 &          .                     & W            \\
        H12 &                   00 38 03.8 &         +37 44 00 &  49.9 &    194.7    & 3.65 &   -396 $\pm$ 10 &     -207 $\pm$ 10 &     -98 $\pm$ 10 &      -23 &          .                     & W            \\
        H14 &                   00 38 49.4 &         +42 22 47 &  18.2 &    327.0    & 1.32 &   -271 $\pm$ 15 &     -76 $\pm$ 15 &      33 $\pm$ 15 &           &          .                     & K            \\
        H15 &                   00 40 13.2 &         +35 52 36 &  74.0 &    185.4    & 5.42 &   -367 $\pm$ 10 &     -182 $\pm$ 10 &     -73 $\pm$ 10 &       -6 &          .                     & W            \\
        H17 &                   00 42 23.6 &         +37 14 34 &  55.0 &    181.0    & 3.97 &   -246 $\pm$ 16 &     -60 $\pm$ 16 &      48 $\pm$ 16 &       122 &          .                     & K            \\
        H18 &                   00 43 36.0 &         +44 58 59 &  50.8 &    2.4      & 3.72 &   -206 $\pm$ 21 &     -10 $\pm$ 21 &      99 $\pm$ 21 &        35 &          .                     & W            \\
        H19 &                   00 44 14.8 &         +38 25 42 &  39.0 &     174.1   & 2.85 &   -272 $\pm$ 18 &     -85 $\pm$ 18 &      24 $\pm$ 18 &        79 &          .                     & W            \\
        H22 &                   00 49 44.6 &         +38 18 37 &  44.4 &     155.0   & 3.25 &   -311 $\pm$ 6 &      -127 $\pm$ 6 &      -17 $\pm$ 6 &        12 &          .                     & W            \\
        H23 &                   00 54 24.9 &         +39 42 55 &  37.0 &    124.0    & 2.71 &   -377 $\pm$ 11 &     -193 $\pm$ 11 &     -84 $\pm$ 11 &     -100 &          Stream D              & W            \\
        H24 &                   00 55 43.9 &         +42 46 15 &  38.8 &     57.0    & 2.96 &   -121 $\pm$ 15 &     66 $\pm$ 15 &       175 $\pm$ 15 &       91 &          Stream C/D (overlap)  & K            \\
        H25 &                   00 59 34.5 &         +44 05 38 &  57.2 &   46.2      & 4.19 &   -204 $\pm$ 14 &     -16 $\pm$ 14 &      93 $\pm$ 14 &         6 &          .                     & W            \\
        H26 &                   00 59 27.4 &         +37 41 30 &  65.8 &    136.6    & 4.81 &   -411 $\pm$ 7 &      -233 $\pm$ 7 &      -124 $\pm$ 7 &     -121 &          Stream C              & W            \\
        H27$^1$ &               01 07 26.3 &         +35 46 48 &  99.9 &   136.7     & 7.31 &   -291 $\pm$ 5 &      -121 $\pm$ 6 &      -12 $\pm$ 6 &        -9 &          .                     & W            \\
        HEC1 &                  00 25 33.8 &         +40 43 38 &  44.9 &    261.9    & 3.28 &   -233 $\pm$ 9 &      -34 $\pm$ 9 &       74 $\pm$ 9 &        145 &          .                     & K, G           \\
        HEC2 &                  00 28 31.5 &         +37 31 23 &  63.5 &    217.4    & 4.64 &   -341 $\pm$ 9 &      -148 $\pm$ 9 &      -39 $\pm$ 9 &        48 &          .                     & G            \\
        HEC6 &                  00 38 35.4 &         +44 16 51 &  42.5 &   346.2     & 3.11 &   -132 $\pm$ 12 &     65 $\pm$ 12 &       174 $\pm$ 12 &      130 &          .                     & G            \\
        HEC10 &                00 54 36.4 &         +44 58 44 &  58.7 &    29.3      & 4.30 &   -98 $\pm$ 5 &       93 $\pm$ 6 &        202 $\pm$ 6 &       119 &          .                     & G            \\
        HEC11 &                 00 55 17.4 &         +38 51 01 &  46.6 &     134.2   & 3.41 &   -215 $\pm$ 5 &      -33 $\pm$ 6 &       76 $\pm$ 6 &         75 &          Stream D              & G            \\
        HEC12$^\star$ &         00 58 15.4 &          +38 03 01 &  60.0 &  135.9     & 4.39 &   -288 $\pm$ 2 &      -109 $\pm$ 4 &      0 $\pm$ 4 &           1 &          Stream C              & .            \\
        HEC13 &                  00 58 17.1 &         +37 13 49 &  68.8 &    142.1   & 5.04 &   -366 $\pm$ 5 &      -188 $\pm$ 6 &      -79 $\pm$ 6 &       -68 &          Stream C              & G            \\
        MGC1$^1$ &              00 50 42.4 &         +32 54 58 &  116.2 &   168.6    & 8.50 &   -355 $\pm$ 7 &      -181 $\pm$ 7 &      -73 $\pm$ 7 &       -23 &          .                     & W, G            \\
        PAndAS-01 &             23 57 12.0 &         +43 33 08 &  118.9 &   289.0    & 8.62 &   -333 $\pm$ 21 &     -119 $\pm$ 21 &     -11 $\pm$ 21 &       28 &          .                     & K            \\
        PAndAS-02 &             23 57 55.6 &         +41 46 49 &  114.7 &  277.2     & 8.40 &   -266 $\pm$ 4 &      -54 $\pm$ 4 &       53 $\pm$ 4 &        108 &          .                     & G            \\
        PAndAS-04 &             00 04 42.9 &         +47 21 42 &  124.6 &  315.1     & 9.12 &   -397 $\pm$ 7 &      -183 $\pm$ 7 &      -75 $\pm$ 7 &       -74 &          NW stream             & W            \\
        PAndAS-05 &             00 05 24.1 &         +43 55 35 &  100.6 &  294.3     & 7.36 &   -183 $\pm$ 7 &      28 $\pm$ 7 &        136 $\pm$ 7 &       168 &          .                     & G            \\
        PAndAS-06 &             00 06 11.9 &         +41 41 20 &  93.7 &     276.5   & 6.75 &   -327 $\pm$ 15 &     -119 $\pm$ 15 &     -10 $\pm$ 15 &       45 &           .                    & K            \\
        PAndAS-07$^2$ &         00 10 51.3 &          +39 35 58 &  86.0 &    257.2   & 6.29 &   -452 $\pm$ 18 &     -248 $\pm$ 18 &     -139 $\pm$ 18 &     -46 &          SW cloud              & G            \\
        PAndAS-08$^2$ &         00 12 52.4 &          +38 17 47 &  88.3 &    245.0   & 6.46 &   -416 $\pm$ 8 &      -215 $\pm$ 8 &      -106 $\pm$ 8 &      -19 &          SW cloud              & G            \\
        PAndAS-09 &             00 12 54.6 &         +45 05 55 &  90.8 &   307.7     & 6.60 &   -444 $\pm$ 21 &     -235 $\pm$ 21 &     -126 $\pm$ 21 &    -115 &          NW stream             & K            \\
        PAndAS-10 &             00 13 38.6 &         +45 11 11 &  90.0 &   308.9     & 6.59 &   -435 $\pm$ 10 &     -226 $\pm$ 10 &     -117 $\pm$ 10 &    -108 &          NW stream             & G            \\
        PAndAS-11 &             00 14 55.6 &         +44 37 16 &  83.2 &   305.7     & 6.09 &   -447 $\pm$ 13 &     -239 $\pm$ 13 &     -130 $\pm$ 13 &    -116 &          NW stream             & W            \\
        PAndAS-12 &             00 17 40.0 &         +43 18 39 &  69.2 &    295.9    & 5.06 &   -472 $\pm$ 5 &      -267 $\pm$ 5 &      -157 $\pm$ 5 &     -129 &          NW stream             & G            \\
        PAndAS-13 &             00 17 42.7 &         +43 04 31 &  68.0 &   293.4     & 4.90 &   -570 $\pm$ 45 &     -365 $\pm$ 45 &     -256 $\pm$ 45 &    -224 &          NW stream             & K            \\
        PAndAS-14 &             00 20 33.8 &         +36 39 34 &  86.2 &    224.9    & 6.31 &   -363 $\pm$ 9 &      -167 $\pm$ 9 &      -58 $\pm$ 9 &        29 &          SW cloud              & W            \\
        PAndAS-15 &             00 22 44.0 &         +41 56 14 &  51.9 &   281.8     & 3.80 &   -385 $\pm$ 6 &      -183 $\pm$ 6 &      -74 $\pm$ 6 &       -26 &          NW stream             & G            \\
        PAndAS-16 &             00 24 59.9 &         +39 42 13 &  50.8 &  246.6      & 3.60 &   -490 $\pm$ 15 &     -292 $\pm$ 15 &     -183 $\pm$ 15 &    -102 &          .                     & K            \\
        PAndAS-17 &             00 26 52.2 &         +38 44 58 &  53.9 &  231.6      & 3.83 &   -279 $\pm$ 15 &     -84 $\pm$ 15 &      25 $\pm$ 15 &       112 &          .                     & K            \\
        PAndAS-18 &             00 28 23.2 &         +39 55 04 &  41.6 &   244.8     & 3.08 &   -551 $\pm$ 18 &     -354 $\pm$ 18 &     -245 $\pm$ 18 &    -163 &          Association 2         & G            \\
        PAndAS-19 &             00 30 12.2 &         +39 50 59 &  37.9 &  240.2      & 2.77 &   -544 $\pm$ 6 &      -348 $\pm$ 6 &      -239 $\pm$ 6 &     -155 &          Association 2         & G            \\
        PAndAS-21 &             00 31 27.5 &         +39 32 21 &  37.7 &  232.1      & 2.76 &   -600 $\pm$ 7 &      -405 $\pm$ 7 &      -296 $\pm$ 7 &     -210 &          Association 2         & W            \\
        PAndAS-22 &             00 32 08.3 &         +40 37 31 &  28.7 &   253.0     & 2.10 &   -437 $\pm$ 1 &      -241 $\pm$ 3 &      -132 $\pm$ 3 &      -55 &          Association 2         & G            \\
        PAndAS-23 &             00 33 14.1 &         +39 35 15 &  33.7 &   227.9     & 2.47 &   -476 $\pm$ 5 &      -282 $\pm$ 6 &      -172 $\pm$ 6 &      -86 &          Association 2         & G            \\
        PAndAS-27 &             00 35 13.5 &         +45 10 37 &  56.6 &    341.3    & 4.14 &   -46 $\pm$ 8 &       154 $\pm$ 8 &       262 $\pm$ 8 &       225 &          .                     & W            \\
        PAndAS-36 &             00 44 45.5 &         +43 26 34 &  30.1 &    9.6      & 2.21 &   -399 $\pm$ 7 &      -205 $\pm$ 7 &      -96 $\pm$ 7 &      -167 &          .                     & W            \\
        PAndAS-37 &             00 48 26.5 &         +37 55 42 &  48.1 &    161.3    & 3.50 &   -404 $\pm$ 15 &     -220 $\pm$ 15 &     -111 $\pm$ 15 &     -72 &          .                     & K            \\
        PAndAS-41 &             00 53 39.5 &         +42 35 14 &  33.1 &    56.1     & 2.42 &   -94 $\pm$ 8 &       94 $\pm$ 8 &        203 $\pm$ 8 &       118 &    Stream C/D (overlap)        & W            \\
        PAndAS-42 &             00 56 38.0 &         +39 40 25 &  42.2 &    120.0    & 3.09 &   -176 $\pm$ 4 &      7 $\pm$ 5 &         115 $\pm$ 5 &        93 &    Stream D                    & G            \\
        PAndAS-43 &             00 56 38.8 &         +42 27 17 &  38.9 &     64.2    & 2.85 &   -135 $\pm$ 6 &      52 $\pm$ 7 &        160 $\pm$ 7 &        79 &    Stream C/D (overlap)        & G            \\
        PAndAS-44 &             00 57 55.8 &         +41 42 57 &  39.4 &    79.8     & 2.99 &   -349 $\pm$ 11 &     -164 $\pm$ 11 &     -54 $\pm$ 11 &     -126 &    Stream C/D (overlap)        & K            \\
        PAndAS-45 &             00 58 37.9 &         +41 57 11 &  41.7 &   75.7      & 3.05 &   -135 $\pm$ 16 &     50 $\pm$ 16 &       159 $\pm$ 16 &       85 &    Stream C/D (overlap)        & G            \\
        PAndAS-46 &             00 58 56.3 &         +42 27 38 &  44.3 &   67.1      & 3.36 &   -132 $\pm$ 16 &     54 $\pm$ 16 &       162 $\pm$ 16 &       82 &    Stream C/D (overlap)        & K            \\
        PAndAS-47 &             00 59 04.7 &         +42 22 35 &  44.3 &   68.7      & 3.35 &   -359 $\pm$ 16 &     -174 $\pm$ 16 &     -64 $\pm$ 16 &     -144 &    Stream C/D (overlap)        & K            \\
        PAndAS-48 &             00 59 28.2 &         +31 29 10 &  141.3 &  159.7     & 10.34 &   -250 $\pm$ 5 &      -83 $\pm$ 6 &       25 $\pm$ 6 &        62 &     .                          & G            \\
        PAndAS-49 &             01 00 50.0 &         +42 18 13 &  48.2 &    71.5     & 3.53 &   -240 $\pm$ 7 &      -55 $\pm$ 7 &       53 $\pm$ 7 &        -24 &    Stream C/D (overlap)        & G            \\
        PAndAS-50 &             01 01 50.6 &         +48 18 19 &  106.7 &  24.1      & 7.81 &   -323 $\pm$ 7 &      -131 $\pm$ 7 &      -22 $\pm$ 7 &       103 &    .                           & G            \\
        PAndAS-51 &             01 02 06.6 &         +42 48 06 &  53.4 &   65.3      & 3.91 &   -226 $\pm$ 5 &      -41 $\pm$ 6 &       67 $\pm$ 6 &        -14 &    .                           & G            \\
        PAndAS-52 &             01 12 47.0 &         +42 25 24 &  78.1 &   75.9      & 5.71 &   -297 $\pm$ 9 &      -118 $\pm$ 9 &      -9 $\pm$ 9 &        -84 &    .                           & W            \\
        PAndAS-53 &             01 17 58.4 &         +39 14 53 &  95.9 &    103.9    & 7.01 &   -253 $\pm$ 10 &     -82 $\pm$ 10 &      26 $\pm$ 10 &       -18 &    .                           & W            \\
       \hline
 \end{tabular}
\end{table*}

\addtocounter{table}{-1}
\begin{table*}
\scriptsize
 \caption{Continued.}
 \begin{tabular}{lccccccccccl}
        Cluster ID              & \multicolumn{2}{c}{Position (J2000.0)} & $R_{\rm proj}$       & PA & $\rho$ & $V_{\rm helio}$  & $V_{\rm gal}$         & $V_{\rm M31corr}$ & $V_{\rm rot-corr}$   & Prominent     & Instrument    \\
                                & RA                &  Dec                                  &  [kpc] & [deg] & [deg] & [km s$^{-1}$] & [km s$^{-1}$] & [km s$^{-1}$]  &   [km s$^{-1}$]             & substructure  &               \\
        \hline
        PAndAS-54 &             01 18 00.1 &         +39 16 59 &  95.8 &     103.6   & 7.01 &   -336 $\pm$ 8 &      -165 $\pm$ 8 &      -56 $\pm$ 8 &       -101 &    .                           & bf            \\
        PAndAS-56 &             01 23 03.5 &         +41 55 11 &  103.3 &    81.7    & 7.56 &   -239 $\pm$ 8 &      -66 $\pm$ 8 &       42 $\pm$ 8 &         -26 &    .                           & bf            \\
        PAndAS-57 &             01 27 47.5 &         +40 40 47 &  116.4 &   90.3     & 8.52 &   -186 $\pm$ 6 &      -18 $\pm$ 7 &       90 $\pm$ 7 &          30 &    Eastern Cloud               & bf            \\
        PAndAS-58 &             01 29 02.1 &         +40 47 08 &  119.4 &  89.4      & 8.74 &   -167 $\pm$ 10 &     1 $\pm$ 10 &        109 $\pm$ 10 &        48 &    Eastern Cloud               & bf            \\
        SK255B &                 00 49 03.0 &         +41 54 57 &  18.4 &   60.8     & 1.34 &   -191 $\pm$ 10 &     -1 $\pm$ 10 &       107 $\pm$ 10 &           &    .                           & bf            \\
        \hline
 \end{tabular}
\end{table*}

\begin{figure*}
\begin{center}
\end{center}
\includegraphics[width = 148mm, angle = 270]{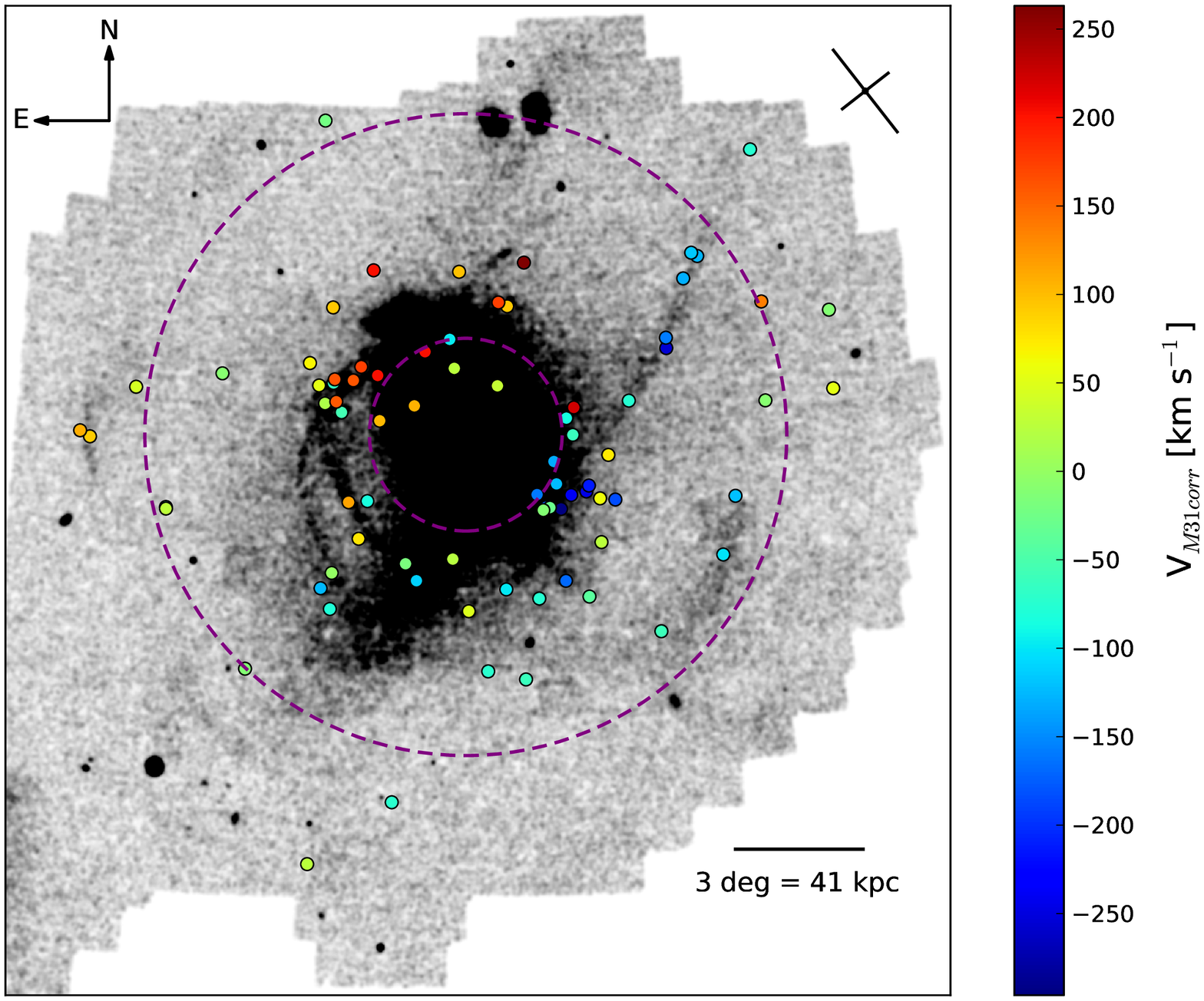}
\caption{The metal-poor ([Fe/H] $\lesssim -1.4$) stellar density map
  of M31 from PAndAS.
  Positions of the observed GCs are marked with coloured dots
  which correspond to their M31-centric radial velocities in
  units of km$\;$s$^{-1}$. As per Section \ref{ss:persp}, the velocities were obtained by correcting for the systemic
  radial motion of M31, which is -109 $\pm$ 4~km$\;$s$^{-1}$ in the Galactocentric
  frame. The purple dashed circles correspond to $R_{\rm proj} = 30$ and 100 kpc. The small schematic in the top right
  shows the orientation of the major and minor axes of M31.}
\label{fig:rvbigmap}
\end{figure*}

\section{Kinematic Analysis}
\label{s:kinematics}
\subsection{The tools: Bayesian inference}
Bayesian inference is a powerful statistical tool for estimating a set of
parameters $\Theta$ in a model $\cal M$, as well as discriminating between
different models. Given some data $D$ and certain prior information $I$,
the mathematical form of Bayes theorem is:
\begin{equation}
p(\Theta | DI) = \frac{p(\Theta | I) p(D| \Theta I)}{p(D|I)}
\label{eq:bayes}
\end{equation}
where $p(\Theta | DI)$ is the posterior probability distribution function (pdf)
of the model parameters, $p(\Theta | I)$ is the prior, and $p(D| \Theta I)$
is the likelihood function. The expression $p(D|I)$ is the Bayesian evidence
which is the average value of the likelihood weighted by the prior and integrated
over the entire parameter space. It is mathematically expressed as:
\begin{equation}
p(D | I) = \int{ p(D| \Theta I) p(\Theta|I)}d\Theta
\label{eq:evidence}
\end{equation}
When more of a model's parameter space has high likelihood values, the evidence
is large. However, the evidence is very small for models with large parameter
spaces having low likelihood values, even if the likelihood function itself is
highly peaked. This Bayesian quantity is key when one wants to discriminate
between two different models, $\cal M$ and $\cal N$. The typical question that
one needs to answer in this case is which model is a better fit to the data.
This can be done via the Bayes factor $B$, which is essentially the ratio between
the evidence of the models that are considered, and formally defined as:
\begin{equation}
B_{\cal{M N}} = \frac{ \int{ p_{\cal{M}}(D| \Theta_{\cal{M}} I_{\cal{M}}) p_{\cal{M}}(\Theta_{\cal{M}}|I_{\cal{M}})}d^m \Theta_{\cal{M}} }{ \int{ p_{\cal{N}}(D| \Theta_{\cal{N}} I_{\cal{N}}) p_{\cal{N}}(\Theta_{\cal{N}}|I_{\cal{N}})}d^n \Theta_{\cal{M}} }
\label{eq:bayesfactor}
\end{equation}
Model $\cal M$ describes the data $D$ better than model $\cal N$ if $B_{\cal MN} > 1$.
A frequently used interpretation scale is the one set up by \citet{Jeffreys61book}, presented
in Table \ref{tab:bayesfactor}.

The Bayes factor is a powerful tool for model selection, especially since
it does not depend on any single set of parameters as the integrations are over the entire
parameter space in each model. This allows for significantly different models to be
compared. In addition, the Bayesian model comparison implicitly guards against
overfitting \citep{Kass95}.

\subsection{Kinematic models}
\label{sec:models}
One of the main goals of this paper is to constrain the overall kinematic properties of the
M31 outer halo GC population. Working in the Bayesian framework provides the ability to discriminate
between different kinematic models, while simultaneously deriving probability distribution functions
for the free parameters in each model.

We construct two kinematic models, ${\cal M}$ and ${\cal N}$. Model ${\cal M}$ comprises
two components: an overall rotation of the M31 outer halo GC system, and the velocity
dispersion of the GCs. Model ${\cal N}$ contains only the velocity dispersion of the GC population.
By considering both a rotating and a non rotating model, we can quantify the statistical
significance of any detected rotation of the M31 outer halo GCs.

The rotation component in ${\cal M}$ is modelled as prescribed in \citet{Cote01}:
\begin{equation}
v_{\rm rot}(\theta) = v_{\rm sys} + A\sin(\theta - \theta_{0})
\label{eq:rot}
\end{equation}
where $v_{\rm rot}$ is the rotational velocity of the GC system at position angle $\theta$,
measured east of north, and $\theta_0$ is the position angle of the rotation axis
of the GC system. The rotation amplitude is labelled as $A$, while $v_{\rm sys}$ denotes the
systemic radial motion of the GC system\footnote{Note that in practice we set this term to be zero,
having already removed the fixed systemic motion of M31 from our GC velocities.}. As detailed
in \citet{Cote01}, this method assumes that the GC system being investigated is spherically
symmetric and that the rotation axis is perpendicular to the line of sight -- i.e., it lies in the
plane of the sky. The model also assumes that the three-dimensional angular velocity is a function
of radial distance only (constant on a sphere). Together these assumptions imply that the projected
angular velocity is a function of projected radius only,  justifying the use of a sinusoid to
describe the rotation of the system.

The velocity dispersion is assumed to have a Gaussian form and to decrease as a function
of projected radius from the M31 centre in a power-law manner. The observed dispersion
($\sigma$) is comprised of two components -- the intrinsic dispersion of the GC system, and the effect
of the measurement uncertainties in the GC radial velocities. This is mathematically
described in Equation \ref{eq:disp}, where $\Delta v$ is the aggregate uncertainty of the
GC velocities, $R$ is the projected radius, and $\gamma$
is the power-law index describing how the velocity dispersion changes
as a function of $R$:
\begin{equation}
\sigma^2  = (\Delta v)^2 + \sigma_0^2 \left(\frac{R}{R_0}\right)^{2 \gamma}
\label{eq:disp}
\end{equation}
The scale radius $R_0$ is fixed at 30 kpc, as this is the point at which the M31
halo begins to dominate \citep[c.f.][]{Geehan06}; $\sigma_0$ is the corresponding intrinsic velocity dispersion
at $R_{\rm proj} = 30$ kpc.

\begin{table}
  \caption{The scale devised by \citet{Jeffreys61book} for discriminating between models -- in this
  case evaluating ${\cal M}$ over ${\cal N}$ -- via the Bayes factor.}
  \begin{tabular}{lll}
    \hline
    \hline
    $\log B_{\cal MN}$	& $B_{\cal MN}$ 	& 	Strength of evidence \\
    \hline
    $<0$ 				& $<1$ 			& 	Negative (supports ${\cal N}$) \\
    0 to 0.5 				& 1 to 3.2 			& 	Barely worth mentioning \\
    0.5 to 1				& 3.2 to 10 		& 	Positive \\
    1 to 1.5				& 10 to 32 		&	Strong \\
    1.5 to 2				& 32 to 100 		& 	Very strong \\
    $>2$ 				& $>100$ 			& 	Decisive  \\
    \hline
  \end{tabular}
  \label{tab:bayesfactor}
\end{table}

Joining Equations \ref{eq:rot} and \ref{eq:disp} we are able to create the rotation enabled
model ${\cal M}$:
\begin{equation}
p_{i,{\cal M}}(v_i, \Delta v_i|v_{\rm rot}, \sigma) = \frac{1}{\sqrt{2\pi\sigma^2}}  \:\: e{^{- \frac{(v_i - v_{\rm rot})^2}{2\sigma^2}}}
\label{eq:rotmodel}
\end{equation}
where $v_{\rm rot}$ is the systemic rotation described by Equation \ref{eq:rot}, $v_i$ are the observed
radial velocities of the GCs as presented in Table \ref{tab:rvtable}, and $\sigma$ is the
velocity dispersion as prescribed in Equation \ref{eq:disp}.

Similarly, the model ${\cal N}$ which does not contain an overall rotation component is simply constructed as:
\begin{equation}
p_{i,{\cal N}}(v_i,\Delta v_i|\sigma) = \frac{1}{\sqrt{2\pi\sigma^2}}  \:\: e^{- \frac{v_i^2}{2\sigma^2}}
\label{eq:norotmodel}
\end{equation}
Note that model ${\cal N}$ is clearly a member of the family of models ${\cal M}$ -- it is the special case where the
amplitude of rotation is zero. In principle, therefore, we could assess the likelihood of this model relative to the favoured
model in the family ${\cal M}$ simply by considering the marginalised probability distributions for the latter.
However, for clarity we prefer to make an explicit comparison between the two models ${\cal M}$ and ${\cal N}$ using
the Bayesian evidence.

Having defined our models, the likelihood function for each of them is:
\begin{equation}
p_{{\cal M}}(D|\Theta) = {\cal L}_{{\cal M}}\left(v, \Delta v, R, \theta | A,\theta_0,\sigma_0,\gamma \right)  = \prod_i p_{i,{\cal M}} 
\label{eq:like1}
\end{equation}
\begin{equation}
p_{{\cal N}}(D|\Theta) = {\cal L}_{{\cal N}}\left(v, \Delta v, R | \sigma_0,\gamma \right)  = \prod_i p_{i,{\cal N}} 
\label{eq:like2}
\end{equation}
in which $v, \Delta v, R, \theta$ are the observed properties of the GCs, and $A,\theta_0,\sigma_0,\gamma$
are the free parameters of the models we are trying to determine. The index $i$ loops
over each individual data point. In all our models we assume flat priors.
Previous studies \citep[e.g,][]{Lee08, Veljanoski13a} have found
the velocity dispersion and the overall rotation of the M31 GCs in both the halo and the disk
to be roughly equivalent in magnitude. Thus, it is important to note that in our proposed
model ${\cal M}$ we are attempting to describe the rotation and velocity dispersion
simultaneously rather than separately as has been the case in the majority of past studies.
This is done in order to avoid any possible bias that can arise from measuring these quantities
in succession, because in such cases the latter measurement depends on the first.

As a reminder, our input sample of GCs is defined by the 72 objects in Table
\ref{tab:rvtable} with $R_{\rm proj} > 30$\ kpc (our 71 observed targets plus HEC12).
The vast majority of velocity measurements for this sample come from our observations as defined in
Section \ref{s:obs}, except in a handful of cases where previous measurements from the literature
are more precise. The spatial coverage of the input sample is high but non-uniform, being slightly
biased towards GCs that (i) project onto visible substructures in the field halo, and (ii) lie at
larger $R_{\rm proj}$.

Calculating the likelihood function, the evidence and the posterior probability distributions
as described in Equations \ref{eq:bayes}, \ref{eq:evidence}, \ref{eq:like1} and \ref{eq:like2}
can be numerically challenging. Various Monte Carlo algorithms
\citep[e.g.][]{Lewis02,Skilling04,Feroz08,Feroz09} have been introduced to make the calculation of
these quantities more efficient. Even though these methods greatly reduce the computation time,
and have been thoroughly tested and widely applied, they do not fully sample the entire parameter
space and there is always danger that a secondary peak in a posterior distribution might remain
undetected, or that the algorithm might get stuck in a local minimum. Because our models contain a
low number of free parameters, we choose to fully sample the parameter space via a brute-force
exploration method. The likelihood function is systematically calculated for each combination of
the free parameters stated in Equations \ref{eq:like1} and \ref{eq:like2}. In this calculation
the amplitude $A$ ranges between 0 and 200 km$\;$s$^{-1}$ with a step size of 3 km$\;$s$^{-1}$, $\theta_0$
ranges between 0 and $2\pi$ radians with an interval of 0.1, $\sigma_0$ ranges between 50
and 600 km$\;$s$^{-1}$ with a 5 km$\;$s$^{-1}$ increment, and $\gamma$ ranges between -1.5 and 0.5 with a step size
of 0.025. Careful testing has shown that this combination of parameters and sampling gives
excellent balance between computational speed and resolution of the likelihood function and the
posterior probability distributions. Finally, the integral in Equation \ref{eq:evidence} is
evaluated by applying the Simpson rule in multiple dimensions.

\begin{table*}
 \caption{ The peak and mean values of the posterior probability distribution functions
 for each free parameter in the two kinematic models, accompanied by their corresponding
 68\% confidence limits. The logarithm of the Bayesian evidence along with the number
 of GCs used for the statistics are also displayed.}
 \label{tab:resall}
\begin{tabular}{ccccccccccc}
 	\hline
	\hline
	Kinematic	& peak $A$		& mean $A$		& peak $\theta_0$	& mean $\theta_0$	& peak $\sigma_0$	& mean $\sigma_0$	& peak $\gamma$			& mean $\gamma$			& log$_{10}(B)$	&$N_{GC}$	\\
	model		& [km$\;$s$^{-1}$]	& [km$\;$s$^{-1}$]	& [deg]			& [deg]			& [km$\;$s$^{-1}$]	& [km$\;$s$^{-1}$]	&				& 				&  		&		\\
	\hline
        \vspace{1.5 mm}
	${\cal M}$	& $86 \pm 17$	        & $86 \pm 17$         	& $135 \pm 11$	        & $135 \pm 11$  	& $129^{+22}_{-24}$	& $136^{+29}_{-20}$	& $-0.45 \pm 0.22$              & $-0.45 \pm 0.22$             	& -191		& 72		\\
	${\cal N}$	& ...			& ...			& ...			& ...			& $209^{+35}_{-38}$	& $222^{+48}_{-32}$	& $-0.37 \pm 0.21$	        & $-0.37 \pm 0.21$              & -218		& 72		\\
	\hline
 \end{tabular}
\end{table*}

\begin{figure*}
\includegraphics[width = 138mm]{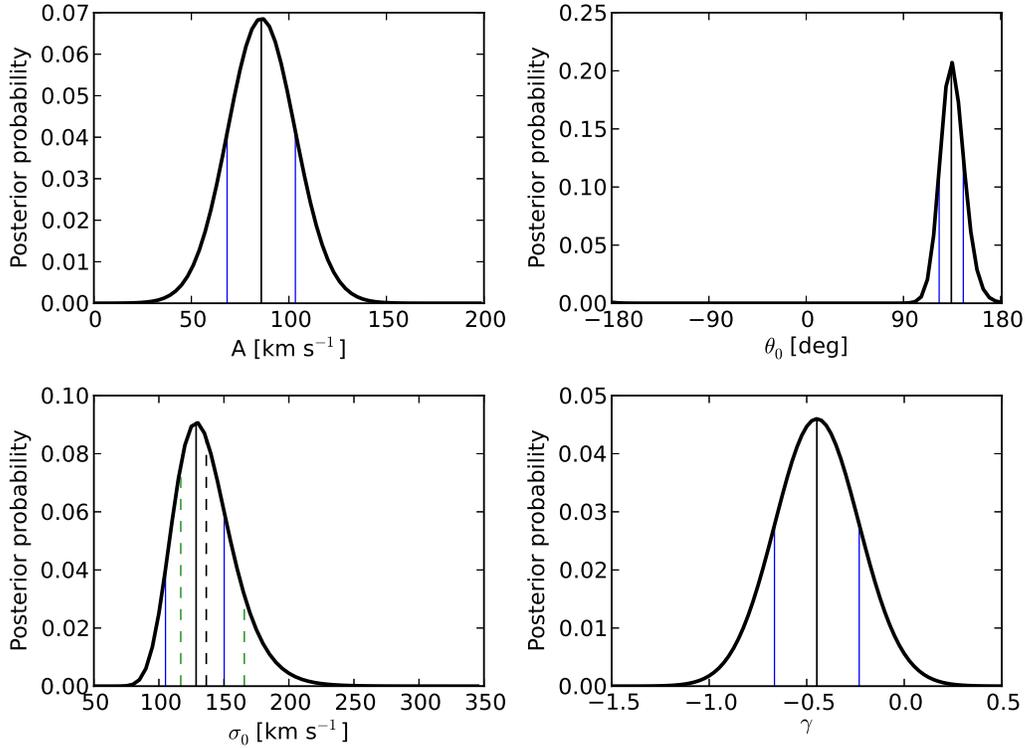}
\caption{Marginalized posterior probability distribution functions for $A$, $\theta_{0}$,
$\sigma_0 $, and $\gamma $ for model ${\cal M}$,
which best represents the observed data. The peak value in each case is marked with a vertical black solid line, while that for the mean,
if different, is marked with the black dashed line. The vertical solid blue lines represent the 1$\sigma$ limits around the peak, while the green
dashed lines mark the 1$\sigma$ limits around the mean, if different from those around the peak. }
\label{fig:pdf}
\end{figure*}

\subsection{Overall halo kinematics}
In \citet{Veljanoski13a} we presented the first
kinematic analysis for a significant number of outer halo M31 GCs.
We discovered that (i) these clusters exhibit substantial net rotation; (ii) they share
the same rotation direction and a similar rotation axis to centrally-located GCs; and
(iii) this rotation axis is approximately coincident with the optical minor axis of M31.
We also observed a hint of decreasing
velocity dispersion with increasing galactocentric radius.
In our present study, we want to determine the statistical significance of
these phenomena by employing the models and methodology presented in the previous subsection.

This allows us to derive posterior probability distribution functions for the free parameters
of each model. Since these distribution functions are not necessarily Gaussian (or even symmetric), we
report both their peak and mean values accompanied by their 68\% confidence limits in Table
\ref{tab:resall}. This table also displays the logarithm of the Bayesian evidence for each model,
which is used to discriminate between them. We find that the rotating model is decisively
preferred over the non rotating one, with log($B_{{\cal MN}}) \approx 27$. The inferred amplitude
of the rotation is $A = 86 \pm 17$ km$\,$s$^{-1}$.

\begin{figure*}
\center
\includegraphics[width = 85mm]{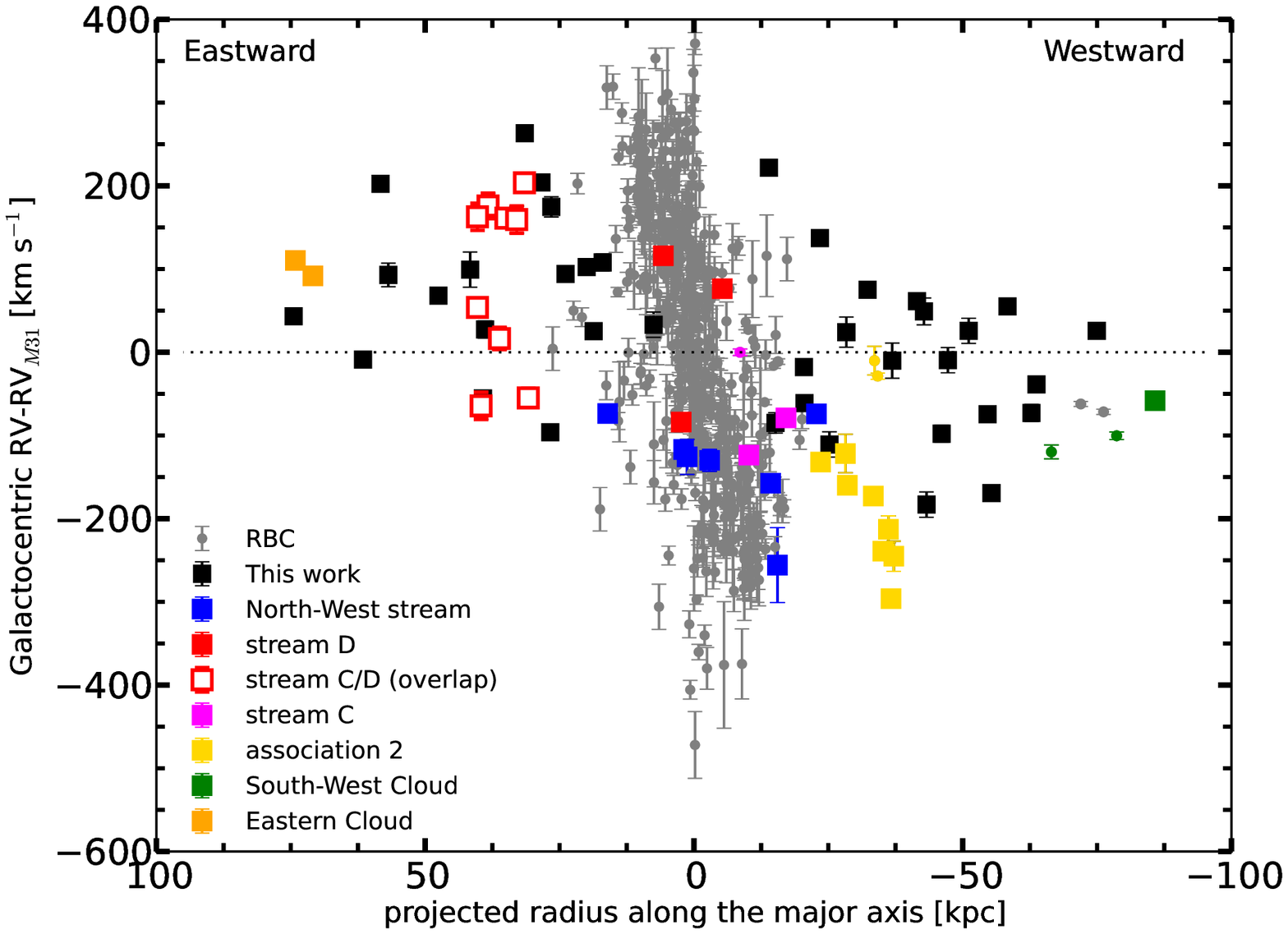}
\includegraphics[width = 85mm]{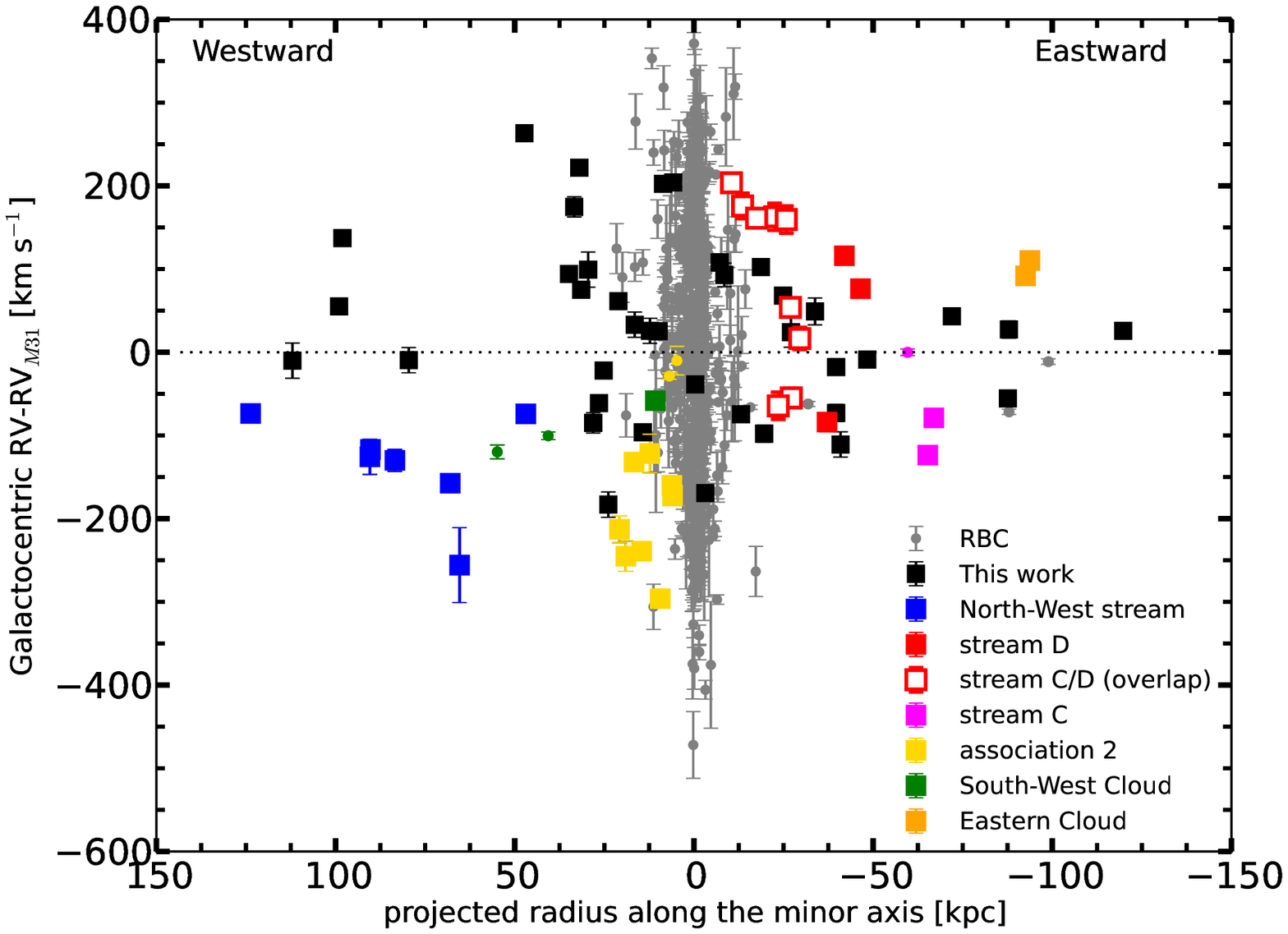}
\caption{Galactocentric velocities, corrected for the systemic motion of M31, versus
 projected distance along the major (left panel) and minor (right panel) optical axes of M31.
 The colours mark GCs that lie on specific stellar debris features, further
 discussed in Section \ref{sec:str}. The left panel clearly displays the rotation of
 the outer halo GCs, which is found to be in the same sense as for their inner region
 counterparts, but with a smaller amplitude. Notice that the rotation is observed even
 for the most distant GCs in projection in our sample. Since there is no clear pattern
 observed in the right panel, this is a good indication that the minor axis is consistent
 with being the rotation axis of GCs located in both the inner regions and the outer halo of
 M31.}
\label{fig:XX}
\end{figure*}

The posterior probability
distribution functions for the free parameters of model ${\cal M}$ are shown in Figure \ref{fig:pdf},
along with the 68\% confidence limits around the peak and mean of each distribution. The strong
preference for the favoured model ${\cal M}$ over the non rotating model ${\cal N}$ (for which $A=0$)
can clearly be seen from the upper left panel. The
position angle of the M31 optical minor axis is 135$^\circ$ east of north, matching
the inferred rotation axis of the M31 outer halo GC system. As expected, the rotation
of the outer GCs is in the same direction as their inner region counterparts
albeit with a smaller amplitude. This is best seen in Figure \ref{fig:XX}, which shows the
Galactocentric radial velocities of the GCs in our sample, corrected for the systemic radial motion of M31,
versus their projected radii along the major (left panel) and minor (right panel) optical axes. The left
panel of Figure \ref{fig:XX} clearly shows that the rotation is observed even for the GCs with the
largest projected distances, and is not driven solely by clusters projected onto major halo substructures
or by clusters not lying on any visible substructure.

When modelling the rotation of the outer halo GC population, we assumed that the rotation axis
lies in the plane of the sky -- i.e., perpendicular to the line of sight. Thus, so far we have determined
the \emph{projected} rotation amplitude, and there is an additional unknown factor $\sin i$ to account
for, where $i$ is the inclination angle of the rotation axis to the plane of the sky. As we have found the
rotation axis of the M31 outer halo GC population to coincide with the minor optical axis of this galaxy, it is
possible that the rotation axis lies perpendicular to the disk of M31. In this case, taking the inclination of M31
with respect to our line of sight to be $77.5\degr$ \citep{Ferguson02}, the peak and mean of the deprojected
rotation amplitude posterior probability distribution function would both be  $88 \pm 17$ km$\,$s$^{-1}$,
barely different from the projected values.

We also find substantial evidence for decreasing velocity dispersion with increasing $R_{\rm proj}$.
Looking at the bottom right panel of Figure \ref{fig:pdf}, it is seen that the peak and the mean of
the $\gamma$ posterior probability distribution function are inconsistent with
$\gamma = 0$. In fact, the posterior probability to measure $\gamma = 0$ is less than 1\%. This is
shown in more detail in Figure \ref{fig:contour}, which shows the 1, 2 and 3$\sigma$ levels of the
likelihood on the $\sigma_0$-$\gamma$ plane. It can easily be seen that a constant velocity dispersion
as a function of $R_{\rm proj}$ may be discarded at approximately the 2$\sigma$ level.

\begin{figure}
\vspace{-6mm}
\includegraphics[width = 85mm, angle = 0]{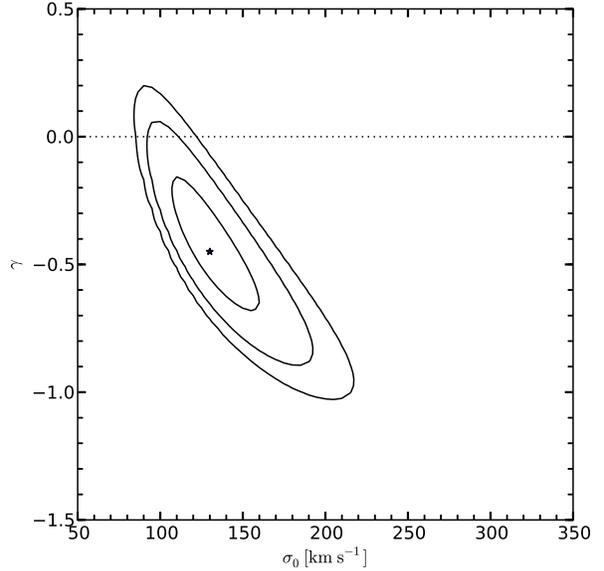}
\caption{Likelihood contours corresponding to the 1, 2 and 3$\sigma$ intervals
 in the $\sigma_0$-$\gamma$ plane. The posterior probability of measuring $\gamma = 0$,
 $p(\gamma = 0) < 1\%$. Thus, a constant velocity dispersion as a function of $R_{\rm proj}$ can
 be almost entirely rejected.}
\label{fig:contour}
\end{figure}

Figure \ref{fig:velproj} shows the Galactocentric radial velocities of the
outer halo GCs, corrected for both their bulk rotation as per model ${\cal M}$ as well as for the
systemic radial motion of M31, as a function of their projected radii (for convenience, we list
the rotation-corrected velocities in Table \ref{tab:rvtable}). Different groups of GCs that lie
along specific stellar streams are marked (see Section \ref{sec:str}). The
GC halo dispersion profile is displayed as a solid line as described by Equation \ref{eq:disp} using
the best fit parameters from Table \ref{tab:resall}. We also plotted the stellar velocity dispersion
profile determined by \citet{Chapman06} for metal-poor giant stars in the range between
$\sim$ 10 and 70 kpc in projection, with the majority of the data points lying between $\sim$ 10
and 50 kpc. The stellar profile was assumed to be linear in shape. Note that beyond 70 kpc we
have used a simple linear extrapolation. Figure \ref{fig:velproj} shows a close similarity between
the velocity dispersions of the M31 halo stars and GCs, despite being fitted by different models,
at least out to $\sim$80 kpc in projection. This similarity might imply that the spatial density
profiles of the M31 halo stars and globulars are also similar, and indeed \citet{Huxor11} have shown
this to be the case by comparing the radial number density profile of the M31 GCs to that of
the metal poor (${\rm  -3.0 < [Fe/H] < -0.7}$) stars (see their Figure 9).

For consistency, we
note that the best-fit parameters for model ${\cal M}$ are in very good agreement with the
results presented in \citet{Veljanoski13a}, where we determined the rotation and velocity
dispersion separately and using only a subset of our currently-available radial velocity sample.

\begin{figure}
\includegraphics[width = 85mm, angle = 0]{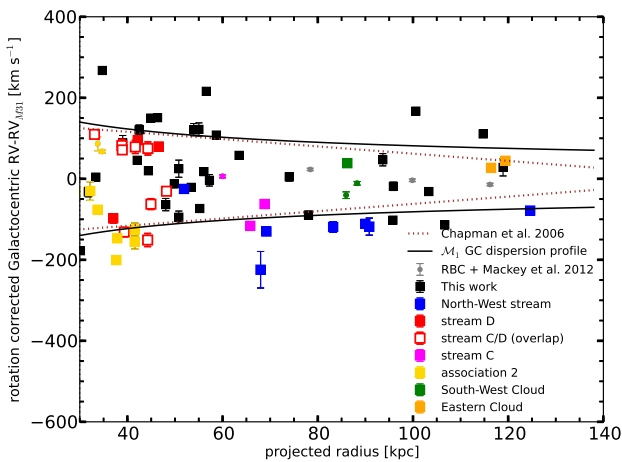}
\caption{Galactocentric radial velocities for our GC sample, corrected for the measured rotation and
 the systemic motion of M31, versus projected radius from the M31 centre. The different coloured
 symbols mark groups of GCs projected on various stellar streams as indicated. The solid line corresponds
 to our most-likely velocity dispersion profile for the outer halo GCs. The dotted line describes the
 velocity dispersion of kinematically selected metal-poor giant stars as measured by \citet{Chapman06}.
 Note that in the latter case, the fit beyond 70 kpc is a linear extrapolation.}
\label{fig:velproj}
\end{figure}

\section{GC groups on streams}
\label{sec:str}
In the previous section we treated the M31 outer halo GCs as a single system and attempted to
describe its global continuous properties. However, the M31 halo is rich with various
field substructures in the form of stellar streams, loops and filaments. Many remote GCs appear
spatially associated in projection with prominent features that are visible in the stellar maps
\citep{Mackey10b}, including a significant number of the clusters in our spectroscopic
sample. Examination of Figure \ref{fig:rvbigmap} reveals that objects projected onto a given feature tend to
exhibit correlated velocities. It might naively be expected that coherent velocity patterns amongst GC groups
would be quite unlikely to arise in the case where the GCs are randomly-selected members of a pressure-supported
halo (even if a substantial rotation component is also present), but would be unsurprising in the case where
they are associated with an underlying kinematically cold stellar debris feature. In the following sub-sections
we consider several GC groups that project onto the main stellar substructures seen in the M31 halo,
as marked on Figure \ref{fig:streammap}, and attempt to indicatively assess the
significance of any observed velocity patterns.

We proceed by employing simple Monte Carlo experiments similar to the one devised by \citet{Mackey13}.
Although these tests are tailored to each specific instance, they all share a common basis.  Our most-likely global
kinematic model from the previous Section tells us what halo velocity dispersion and systemic rotation to expect at each
GC position. For the $N$ GCs in a given group, we first subtract the global rotation signal from the observed velocities $V_{\rm M31corr}$
(which have, of course, already been corrected for the M31 systemic motion)\footnote{We remind the reader that these
rotation-corrected GC velocities are listed in Table \ref{tab:rvtable} for easy reference.}, and then generate 10$^6$ sets of
$N$ GCs with positions matching those of the real set, but with each individual velocity randomly drawn from a Gaussian distribution
centred on zero and with a width set by the dispersion model described by Equation \ref{eq:disp} at the appropriate projected radius.
This shows us what typical velocity configuration(s) to expect for the null hypothesis that all $N$ GCs are independent,
uncorrelated members of the M31 halo, and thus allows us to broadly quantify how unusual any observed velocity pattern
might be in this context.

Note that, ideally, we would simply match our GC velocities to kinematic measurements derived directly from a given
stellar substructure to establish or refute any association between them. However, determining velocities for these extremely
low surface-brightness features is a challenging observational problem and at present few such measurements
have been published. In what follows we highlight only a couple of cases where it is possible to make such a direct comparison using
extant data.

\begin{figure*}
\includegraphics[width=148mm, angle = 270]{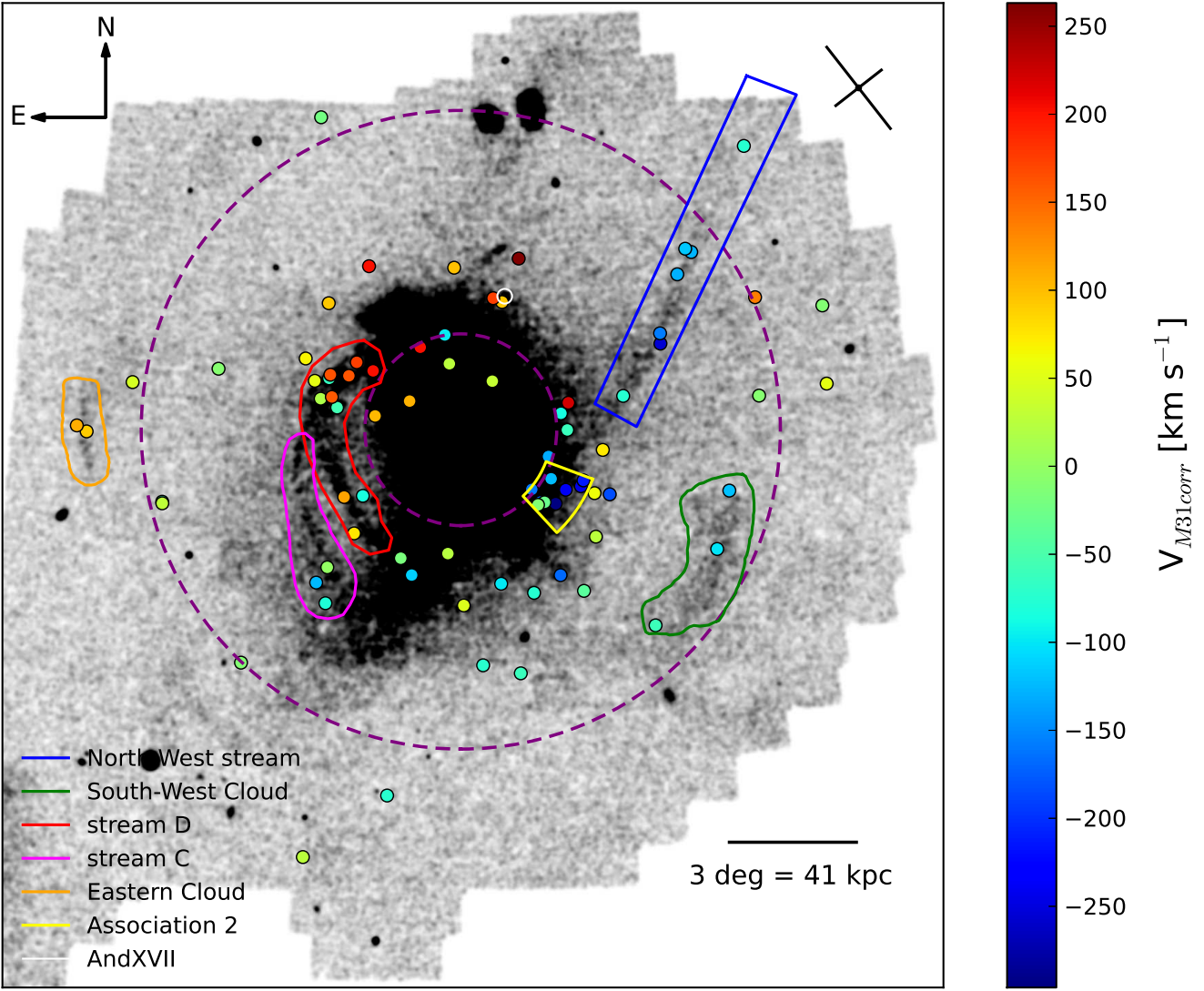}
\caption{The metal-poor stellar density map of M31 from PAndAS.
Points are as in Figure \ref{fig:rvbigmap}. Coloured contours mark
the cold stellar features of interest. The white circle marks the
position of the And XVII dSph (see text for details). }
\label{fig:streammap}
\end{figure*}

\subsection{The North-West Stream}
\label{ss:nws}
The North-West Stream is a narrow stellar debris feature extending radially over a range $R_{\rm proj} \sim 30 - 130$ kpc.
Projected on top lie 7 GCs for which we have measured velocities. Because of the radial nature of the stream, it is
interesting to examine how the velocities of these GCs behave as a function of $R_{\rm proj}$. This is shown in Figure
\ref{fig:NWs}, where we plot Galactocentric velocity, corrected for the measured rotation and systemic motion of M31,
against projected radius.

Six of the seven NW Stream GCs share a clear trend in corrected radial velocity as a function of
$R_{\rm proj}$, in that the velocity becomes more strongly negative the closer a GC is to the
centre of M31. However the innermost GC, PA-15 (the spectrum for which has S/N $\approx 8$ per \AA),
deviates substantially from this trend and, assuming its measured velocity is correct, it is
difficult to see how this object could be associated with the NW Stream despite the fact that its
position projects precisely onto the feature.

As marked in Figure \ref{fig:NWs}, the relationship between the outermost five GCs on the stream is very close to
linear, with a gradient of $1.0 \pm 0.1$ km$\,$s$^{-1}$ per kpc, a zero-point of $-199 \pm 9$ km$\,$s$^{-1}$, and a
Pearson correlation coefficient ${\cal R} = 0.98$. This is remarkable -- we do not know of any compelling reason to
expect a highly linear correlation between velocity and radial distance. Indeed we ascribe no important physical insight
into this specific form of the relationship -- fitting a straight line to the data is merely the simplest means of quantifying
the observed trend. It is also notable that the NW Stream clusters lie substantially displaced by a magnitude $\ga 100$ km$\,$s$^{-1}$
from zero velocity, which is where the mean of the distribution of corrected halo velocities should sit. This is larger
than the dispersion of the GC system at commensurate radii (see Figure \ref{fig:velproj}).

\begin{figure}
\includegraphics[width = 85mm, angle = 0]{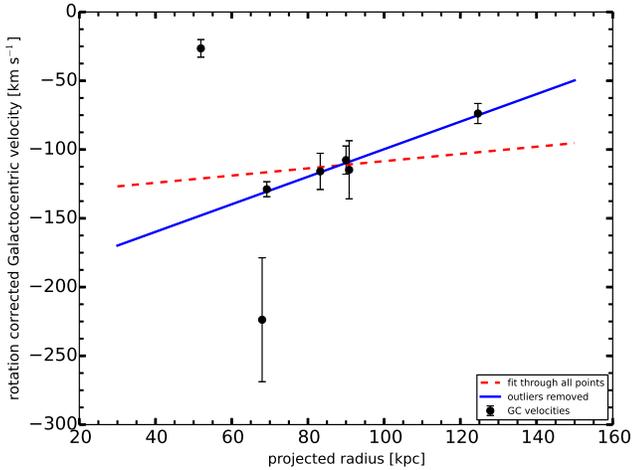}
\caption{Galactocentric radial velocity, corrected for the measured rotation and systemic motion of M31, as a function
of projected radius for the 7 GCs that lie projected on top of the North-West Stream. The dashed red line is a linear fit
through all the data points, while the solid blue line is the fit after excluding the two innermost GCs. This latter fit has
a slope of $1.0 \pm 0.1$ km$\,$s$^{-1}$ per kpc and a zero-point of $-199 \pm 9$ km$\,$s$^{-1}$.}
\label{fig:NWs}
\end{figure}

Although the sixth NW Stream cluster, PA-13, does share in the trend of increasingly negative
velocity with smaller $R_{\rm proj}$, it is quite displaced from the linear relationship described
above. However, this object has the lowest S/N spectrum in our entire GC sample
(S/N $\approx 2$ per \AA), and its velocity is thus accompanied by a large uncertainty
such that its relationship to the stream is ambiguous. A more precise measurement for PA-13 would
clearly be valuable.

We conducted a Monte Carlo experiment to consider the outermost five GCs on the NW Stream. We (conservatively)
counted what fraction of our mock GC sets satisfied ${\cal R} < -0.9$ or ${\cal R} > 0.9$, which is considered an indicator of high
(anti)correlation. Around 3\% of the simulated sets satisfy this criterion. If we only consider cases of infall, meaning the
sets only need to satisfy ${\cal R} > 0.9$, the probability of finding such a pattern falls to 2\%. We also counted
how many times all five GCs fell outside either $\pm 1\sigma$ from the mean (i.e., outside the measured dispersion at given radius).
This is a very unusual configuration, occurring only $0.02$\% of the time. These two simple tests show that the kinematic properties
of the five NW Stream GCs are almost certainly not due to a chance occurrence among independent halo GCs, providing
additional convincing evidence, beyond their spatial alignment, for an association with each other and the underlying stellar stream.
The observed velocity gradient amongst the GCs likely represents the infall trajectory of the progenitor satellite. The scatter
of the five GC velocities about the best-fit line is very small, suggestive of a dynamically cold system with a dispersion consistent with zero.

\subsection{The South-West Cloud}
The South-West Cloud is a large diffuse overdensity near the major axis of M31 at $R_{\rm proj} \sim 90$ kpc. It has been
studied in detail by \citet{Bate14} \citep[see also][]{Lewis13}. There are three GCs projected onto the Cloud, two of which
(PA-7, PA-8) were shown by \citet{Mackey13} to have velocities consistent with being members of this substructure.

Here we have measured a velocity for the third possible member of the sub-group, PA-14, as well as confirming the
velocities for PA-7 and PA-8 from \citet{Mackey13}. Since the SW Cloud closely resembles an arc tracing roughly
constant $R_{\rm proj}$, we consider the rotation-corrected velocities of these three GCs as a function
of position angle relative to the M31 centre (Figure \ref{fig:SWC}). There is a clear velocity gradient along the arc,
in that the corrected velocities become more negative with increasing position angle (i.e., in the anti-clockwise direction
on Figure \ref{fig:streammap}, or from south to north along the arc of the stream).

Once again, a linear fit does an excellent job of describing this trend. The best-fit line has a
gradient of $-2.32 \pm 0.02$ km$\,$s$^{-1}$ per degree, a zero-point of $550 \pm 6$ km$\,$s$^{-1}$,
and a correlation coefficient ${\cal R} = -0.99$. As before, we do not ascribe
any particular significance to this assumed form for the relationship -- a linear fit is just the
simplest means of quantifying the observed trend in velocity with position angle.

\begin{figure}
\includegraphics[width = 85mm, angle = 0]{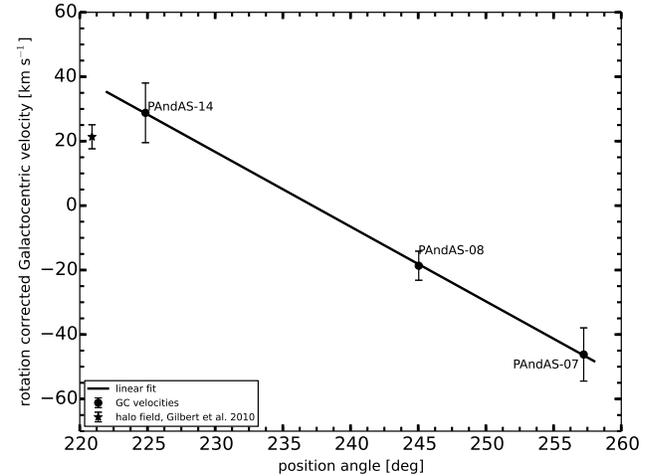}
\caption{Galactocentric radial velocity, corrected for the measured rotation and systemic motion of M31,
as a function of position angle (east of north) relative to the galaxy's centre, for the 3 GCs that lie projected
on top of the South-West Cloud. The solid line represents the best fit, having a gradient of $-2.32 \pm 0.02$
km$\,$s$^{-1}$ per degree and a zero-point of $550 \pm 6$ km$\,$s$^{-1}$. The halo field measured by
\citet{Gilbert12}, in which they detected a cold kinematic peak with $V_{\rm helio} = -373 \pm 3$ km$\,$s$^{-1}$
and an intrinsic dispersion $\sigma = 6.1^{+2.7}_{-1.7}$ km$\,$s$^{-1}$ is also marked.}
\label{fig:SWC}
\end{figure}

We ran a Monte Carlo experiment to consider the three SW Cloud GCs, and assessed what fraction of our mock
sets exhibit a linear correlation with ${\cal R} > 0.9$ or ${\cal R} < -0.9$. This is moderately common, arising $29$\% of
the time. Our calculation supersedes that of \citet{Mackey13}, as their model did not include any correction for the
systemic rotation because its existence was not known at that time. However, since the SW Cloud lies near the
M31 major axis, it is important to account for the rotation signal.

This result is, alone, insufficient to robustly associate this sub-group of three GCs with each other; in addition,
these objects do not have rotation-corrected velocities offset far from the expected mean of zero.
However, we recall that the chance of all three GCs being projected directly onto the SW Cloud in the first place is quite
small \citep[$\sim 2.5$\%,][]{Mackey10b}. Furthermore, \citet{Bate14} have noted that spectroscopic measurements
of the M31 field halo near to PA-14 by \citet{Gilbert12} revealed a cold kinematic peak at a very similar velocity to that
of the cluster: $+21 \pm 4$ km$\,$s$^{-1}$ in the corrected frame, with an intrinsic dispersion $\sigma = 6.1^{+2.7}_{-1.7}$ km$\,$s$^{-1}$
(see Figure \ref{fig:SWC})\footnote{Here we have assumed that the M31 field halo is a subject to the same rotation
effects as the GCs projected onto the SW Cloud, which is not unreasonable if the clusters trace the motions of
the underlying substructure.}. We conclude that, on balance, PA-7, PA-8, and PA-14 are all likely associated with each other and
the SW Cloud.
A radial velocity gradient along the arc, as suggested by the GCs, may imply substantial motion in the plane of the sky
and hence a significant line-of-sight depth to the feature -- as tentatively detected by both \citet{Mackey13} and \citet{Bate14}.

\subsection{Streams C and D}
Streams C and D are two well-defined arc-like substructures due east of M31 \citep[see][]{Ibata07,Richardson11}.
This is a complex part of the M31 halo -- the streams appear to overlap, in projection, at their
northern end; in addition, Stream C is known to split into two kinematically distinct constituents \citep{Chapman08} --
a metal-poor component, Cp, which is the narrow stream visible in Figures \ref{fig:rvbigmap} and \ref{fig:streammap},
and a metal-rich component, Cr, which overlaps Cp in projection but is spatially much broader. This
latter feature is not evident in Figures \ref{fig:rvbigmap} and \ref{fig:streammap} because its member stars
fall above the metal-rich cut-off used in the construction of these plots; see however maps in, e.g., \citet{Martin13,Ibata14}.

There is a total of 15 GCs projected on top of Streams C and D, all of which we have measured velocities for.
The northern area of overlap is particularly rich in clusters, with 9 contained inside a relatively small region on
the sky. Our observations suggest that these GCs split into two distinct kinematic subgroups. We employed the
biweight location and scale estimators \citep{Beers90} to determine the mean velocity and velocity dispersion of each.
The first contains five GCs (H24, PA-41, PA-43, PA-45, PA-46), has a mean rotation-corrected velocity of
$84 \pm 4$ km$\;$s$^{-1}$ and dispersion of $8^{+15}_{-8}$ km$\;$s$^{-1}$, while the second features three
GCs (B517, PA-44, PA-47) with a mean rotation-corrected velocity of $-111 \pm 49$ km$\;$s$^{-1}$ and a dispersion
of $39^{+54}_{-39}$ km$\;$s$^{-1}$. The ninth cluster in the region, PA-49 cannot be robustly identified with either
kinematic subgroup. These measurements supersede those from our earlier work \citep{Veljanoski13a}, where we
identified two similar kinematic groupings, but with an incomplete data set.

We conducted a Monte Carlo experiment for each GC subgroup to assess the likelihood that they are chance
assemblies of unrelated objects. At $R_{\rm proj} \approx 40$ kpc the expected velocity dispersion of halo GCs is
$\sigma \sim 115$ km$\;$s$^{-1}$. For the first subgroup, the fraction of mock sets where all five members lie
outside either $\pm0.7\sigma$ but with an internal velocity dispersion less than 10 km$\;$s$^{-1}$, is extremely
small at 0.02\%. For the second subgroup, the configuration where three GCs all sit outside either
$\pm0.9\sigma$ while having an internal dispersion below 40 km$\;$s$^{-1}$ is somewhat more frequent, occurring
$2\%$ of the time. Combined with the low probability of so many GCs clustering spatially \citep[see][]{Mackey10b},
we conclude that our GC groups are very likely associated with two of the underlying substructures. It is notable
that the mean velocities of the subgroups fall either side of zero. Thus they, and, in all likelihood, two of the three
overlapping streams, are in counter-rotating orbits about M31.

Following Streams C and D anti-clockwise in Figure \ref{fig:streammap}, both have three GCs projected onto
their southern regions. The three clusters projecting onto Stream D have velocities encompassing a range of
$\approx 200$ km$\;$s$^{-1}$; thus these objects do not form a kinematic subgroup. At present we are
unable to robustly assess which, if any, of these GCs might be associated with Stream D.

The three Stream C clusters also have quite disparate velocities and do not constitute a kinematic subgroup.
One of these objects, HEC12 (also known as EC4), is quite well studied. It lies precisely on the narrow metal-poor
component Cp, and shares a common velocity with this stream \citep{Chapman08, Collins09}. Of the other two
GCs, HEC13 lies less than a degree away and has a velocity very similar to that measured for the metal-rich
component of Stream C by \citet{Chapman08} ($V_{\rm helio} = -366 \pm 5$ km$\;$s$^{-1}$ and $-350 \pm 3$ km$\;$s$^{-1}$,
respectively) -- i.e., HEC13 is plausibly associated with Cr. The velocity of the third GC, H26, indicates that it
is most likely a chance alignment that is not a member of either component of Stream C.

\subsection{The Eastern Cloud}
The Eastern Cloud is a small arc-like stellar debris feature located at $R_{\rm proj} \sim118$ kpc due
east of M31. There are two GCs projected onto this overdensity -- PA-57 and PA-58. These have a velocity difference
of less than 20 km$\;$s$^{-1}$ and a mean rotation-corrected velocity that sits $\approx 0.5\sigma$ away from the
expected systemic mean of zero (as before, $\sigma$ is the dispersion of halo GCs at the appropriate radius).

The explore how commonly such an arrangement might occur we conducted a Monte Carlo experiment for
these two GCs. The fraction of mock pairs that, conservatively, have both members sitting outside $\pm 0.5\sigma$
but with an absolute velocity difference between them of smaller than 30 km$\;$s$^{-1}$, is around $10\%$.
However, the Eastern Cloud is a comparatively small overdensity in terms of its projected surface area.
\citet{Mackey10b} did not consider this feature as it had not been discovered at the time of their work.
The arc subtends a position angle of $\approx 15\degr$ and spans, generously, the radial range $115-120$ kpc
($\approx 8.4\degr - 8.8\degr$). Its projected surface area is hence $\approx 0.9$ deg$^{2}$.  We also know that
the surface density of GCs at this radius in the M31 halo is very close to $0.1$ deg$^{-2}$ (Mackey et al. 2014, in prep.).
Applying simple Poisson statistics, we infer that the probability of observing two or more GCs falling in this region
by chance is $\sim 0.4$\%. The chance that they also have very similar velocities, as per the calculation above,
is then $\sim 0.04$\%. We conclude that the two GCs projected onto the Eastern Cloud are almost certainly
associated with this substructure.

\subsection{Association 2}
\citet{Mackey10b} identified a spatial overdensity of GCs, dubbed ``Association 2", close to the
western major axis of M31 at a distance of $R_{\rm proj} \sim 35$ kpc.  It consists of 10
clusters\footnote{Two additional GCs were discovered in this region since the analysis by
\citet{Mackey10b}.} sitting within a small area, constituting the single highest local density
enhancement of GCs, relative to the azimuthal average, seen in the M31 halo. This is a complex
region where the outer disk and stellar halo of M31 overlap, and it is difficult to assess whether
there might be one or more distinct underlying stellar features (see Figure \ref{fig:rvbigmap}).

We have measured velocities for all 10 GCs lying within the Association 2 region. The ensemble
splits naturally into two kinematic sub-groups: (i) a set of four clusters (H2, PA-18, PA-19, PA-21)
for which the biweight estimators give a mean rotation-corrected velocity of
$-162 \pm 18$ km$\;$s$^{-1}$ and a dispersion of $30 \pm 28$ km$\;$s$^{-1}$; and (ii) a second set
of four objects (H7, H8, PA-22, PA-23) for which the biweight indicators suggest a mean
rotation-corrected velocity of $-63 \pm 17$ km$\;$s$^{-1}$ and a dispersion of
$19 \pm 13$ km$\;$s$^{-1}$. The two remaining clusters, G1 and G2, cannot be robustly associated
with either ensemble. Intriguingly, our first GC subgroup consists only of objects with
$R_{\rm proj}$ in the range $\approx 38$-$42$\ kpc, while the second consists only of GCs with
$29 \la R_{\rm proj} \la 34$ kpc. That is, splitting the overall ensemble by velocity, as we have
done, naturally also results in clustering by position.

To assess the plausibility of these two apparently coherent sub-units forming by chance we conducted
a Monte Carlo experiment for each. We find that the occurrence of 4 random GCs all lying outside
either $\pm 1.1\sigma$ but having an internal dispersion less than $30$ km$\;$s$^{-1}$, as per our
first observed subgroup, is very rare at $0.04$\%. Our second subgroup, for which all 4 members sit
outside either $\pm 0.4\sigma$ but have an inter-group dispersion smaller than $20$ km$\;$s$^{-1}$
is also very rare and arises $0.05$\% of the time in our model.

In summary, Association 2 is not a single kinematically coherent ensemble of clusters. Instead it
is primarily comprised of two clearly distinct sub-groups, and thus may possibly represent the
projected convergence of two relic systems. In this context it is interesting that Association 2
sits very close to the expected base of the North-West Stream. Our first GC sub-group, which has
$R_{\rm proj} \sim 40$\ kpc and a rotation-corrected velocity of $\approx -162$ km$\;$s$^{-1}$,
matches very closely to an extrapolation of the linear velocity gradient along the NW Stream that
we described above (see Figure \ref{fig:NWs}), and it is tempting to speculate that this sub-group
may be linked to that feature. Similarly, \citet{Ibata05} studied a discrete fragment of the M31
outer disk very nearby to this region, and found its velocity to sit near
$V_{\rm helio} \approx -450$ km$\;$s$^{-1}$ \citep[see also][]{Reitzel04,Faria07}. This matches
well with the heliocentric velocities of the GCs in our second sub-group
(see Table \ref{tab:rvtable}) -- the weighted mean for which is $\approx -441$ km$\;$s$^{-1}$. It
is therefore possible that these GCs may be associated with the outer disk of M31.

It it also worth nothing that the GC G1 (also known as Mayall II) lies in the Association 2 region.
This cluster is probably the brightest and most massive member of the M31 GC halo system, and it
exhibits a number of peculiar properties analogous to the Galactic GC $\omega$-Cen
\citep[e.g.][]{Ma07,Ma09}, which make it a likely galaxy remnant candidate \citep[e.g.][]{Meylan01}.
It is therefore perhaps surprising that this cluster does not belong in either of the
two kinematically identified groups within Association 2; in particular it does not fit with the trend
set by the GCs associated with the NW Stream.

\subsection{And XVII}
\citet{Irwin08}, in their discovery paper for the M31 dwarf spheroidal (dSph) satellite And XVII, noted that
three GCs lie very close to this system -- H11, HEC6, and HEC3, with projected distances of 2.0 kpc, 3.7 kpc,
and 5.9 kpc respectively. This is the only known instance of a possible association between GCs and an M31
dSph satellite.  We have obtained velocity measurements for both H11 and HEC6, allowing us to assess
whether either of these GCs might plausibly be gravitationally bound to And XVII.

The heliocentric radial velocity of And XVII is $-251 \pm 2$ km$\;$s$^{-1}$ \citep{Collins13} while that for
HEC6 is $-132 \pm 12$ km$\;$s$^{-1}$. This GC, the most distant of the three candidates, is clearly not
associated with the dwarf.  The situation is more complex for H11, which is the closest candidate.
Its velocity is only separated from that of And XVII by $38 \pm 8$ km$\;$s$^{-1}$ in the heliocentric
frame, and hence there is a higher chance it might be gravitationally bound. We investigate this via a
simple Newtonian escape velocity argument:
\begin{equation}
v_{\rm esc} = \sqrt{\frac{2GM_{\rm tot}}{R}}
\label{eq:esc}
\end{equation}
where $v_{\rm esc}$ is the escape velocity, $G$ is the gravitational constant, $M_{\rm tot}$ is the
total mass of the system and $R$ is the distance between the two objects. The only available
mass estimate of And XVII comes from \citet{Collins14}, who find $0.13 \times 10^{7} M_{\odot}$
within the half-light radius of the galaxy. Under the assumption that $R \approx 2$ kpc and
$M_{\rm tot} \approx 10^{7} M_{\odot}$ the escape velocity is found to be
just $\sim 7$ km$\;$s$^{-1}$. Applying the same principles, And XVII would be required to have a
total mass of at least $\sim 3\times 10^{8} M_{\odot}$ in order for H11 to be gravitationally bound.

\section{Discussion}

\subsection{GC Kinematics}
Our high quality PAndAS data has made exploring the true outer halo of M31 ($30 \lesssim R_{\rm proj} \lesssim 150$ kpc)
in a continuous and complete fashion possible for the first time. This region is seen to be dominated
by various stellar debris features, thought to be the remnants of accreted dwarf galaxies \citep[e.g.][]{McConnachie09,Ibata14}.
In addition, a significant portion of the GCs discovered in the outer halo appear to preferentially lie projected on top of these debris
features \citep{Mackey10b}.

Using the measurements presented in the current paper, we have demonstrated that various discrete groups of such GCs
-- specifically those projecting onto the most luminous halo streams and overdensities -- exhibit clear kinematic patterns.
In Section \ref{sec:str} we used our global kinematic measurements from Section \ref{s:kinematics}, in conjunction with
simple Monte Carlo experiments, to indicatively assess how frequently these velocity trends or correlations might occur in
the case where all the GCs in a given group are independent members of the M31 halo. Each instance we examined
(apart from the ostensible And XVII association) showed clear evidence for non-random behaviour, indicative of a dynamical
link between the GCs and the streams or overdensities that they project onto. Together these results strongly reinforce the
notion from \citet{Mackey10b} that a substantial fraction of the outer halo GC population of M31 consists of objects accreted
along with their now-defunct host galaxies.  A striking feature of many of the ensembles we considered is the coldness of
their kinematics -- cluster groups on the NW Stream, SW Cloud, and Eastern Cloud, as well as sub-groups in the Stream
C/D overlap area and in Association 2, all exhibit velocity dispersions consistent with zero. At present, measurements
of stream velocities directly from the constituent stars are available only in a handful of cases; however we have demonstrated
that these few instances largely support our assertions.

In light of these results, our discovery of the high overall degree of coherent rotation exhibited by the outer halo
GC population in M31 is rather surprising. It is relevant to note that this rotation is {\it not} predominantly driven by either the
subset of remote clusters clearly associated with underlying stellar streams, or the complement of this subset.
This is clearly evident from inspection of the left-hand panel of Figure \ref{fig:XX}. We have also found that the outer halo
GCs share the same rotation axis as the inner halo GCs; indeed, the rotation of these
two subsystems is virtually indistinguishable save for the difference in amplitude. This is in contrast with observations in
the Milky Way where the halo GC population appears to exhibit at most only a mild net rotation \citep{Harris01,GCReview06,Deason11}.

It is unfortunate that to date there is little opportunity to compare the kinematics of the M31 outer halo GCs to those of
other massive spiral galaxies apart from the Milky Way. \citet{Olsen04} derived kinematic properties for 6 spiral galaxies
in the Sculptor group. However, the GCs in these galaxies mainly lie in the inner regions of their hosts, and the results
of their study are likely to be affected by small number statistics. \citet{Nantais10} presented a discovery of rapid rotation
in the GC system of M81. However, all of the GCs with available radial velocity data in that galaxy lie at projected
radii of less than 20 kpc. In elliptical galaxies it is common to find rotating GC populations towards more central
regions, but finding significant rotation beyond a few tens of kpc appears to be a rare occurrence \citep[e.g.][]{Woodley10,Strader11,Blom12,Pota13}.

It is interesting to consider how to reconcile our discovery of significant rotation in the outer halo of M31 with the
chaotic accretion of dwarf galaxies implied by hierarchical models. One way this phenomenon might arise is
through the major merger of two spiral galaxies. For example, the numerical simulations of \citet{Bekki10} suggest
that a major merger between M31 and a similar spiral galaxy could give rise to the rapid rotation observed in the
inner GC system of M31, including the rotation of the halo populations. More generally, a large
fraction of the halo GC subsystem might have been brought into the potential well of M31 via a single moderate-mass
satellite. In this event, the satellite seeds its GCs in the halo as it spirals in towards the centre of M31.

This idea
is supported by both the thick disk of M31, which is found to rotate in the same sense (although more rapidly than)
the outer halo GC population \citep{Collins11}, and the kinematic properties of the inner spheroid, which also exhibits
substantial rotation \citep{Dorman}. However, in order for such a satellite to be able to deposit several tens
of GCs it would need to have a considerable mass -- perhaps akin to the Large Magellanic Cloud, which possesses
16 old GCs \citep[e.g.,][]{Mackey06}. If an encounter between M31 and such a massive satellite did occur,
the question must arise as to how disruptive such an event would have been on the M31 disk. In addition, in this
scenario (and indeed that involving the merger of two spirals) it may also be difficult to explain the observed
presence of distinct dynamically cold subgroups of GCs as well as the typically narrow stellar debris streams
in the halo. Detailed modelling is required to address these uncertainties.

Another possibility is that the outer halo GC system of M31 is indeed the product of the assimilation of multiple
dwarf galaxies, but that these were accreted into the M31 potential well from a preferred direction on the sky.
This would be consistent with the observation that many of the surviving dwarf galaxies associated with M31 lie
in a thin rotating planar structure, as reported by \citet{Ibata13}. It is interesting to note that this plane of dwarfs
also appears to be rotating in the same direction as the outer halo GC system, although its rotation axis is
inclined by $\sim 45\degr$ with respect to the minor axis of M31. A similar plane of dwarfs is observed in the
Milky Way \citep[e.g.,][]{Metz07}, and it has been shown that some of the GCs in the outer Galactic halo share
this planar alignment \citep{Keller12,Pawlowski12}. It has been hypothesised that the formation of these planes of dwarfs,
and by extension the possibility of accretion from a preferred direction, could occur as dwarf galaxies move along
large scale dark matter filaments or sheets, in which case they are expected to have aligned angular momenta
as seen in some recent cosmological simulations \citep{Libeskind05,Libeskind11,Lovell11}. An alternative hypothesis
is that these are tidal dwarfs formed during an early gas-rich merger \citep[e.g.,][]{Pawlowski13}.

It is important to note that we are {\it not} suggesting that the properties of the outer halo GC system of M31 stem
directly from the rotating plane of satellites reported by \citet{Ibata13}; indeed, there are almost no GCs with
$R_{\rm proj} > 30$\ kpc lying close to this plane. Nonetheless, the observed fact that a substantial number of dwarf
satellites of M31 possess correlated angular momenta raises the possibility that the parent galaxies of the accreted
outer halo GC population may once have shared a similar, but now disrupted, configuration. In this context it is
relevant that almost all of the dwarf galaxies thought to be members of the planes presently observed in both M31
and the Milky Way are insufficiently massive to host GCs. Hence the outer halo GC population in M31 might still
have been assembled from only a few larger host systems, even if their accretion was related to a previous planar structure.

\subsection{Application: the mass of M31}
An accurate measurement of the mass of M31 is important for constraining the dynamics of Local
Group galaxies, and for testing various cosmological models and predictions. Even though M31
is the closest massive galaxy to our own, it is striking that we have yet been unable to measure its
mass to high precision; indeed there is even still debate as to whether M31 or the Milky Way is
more massive. A number of studies have employed a variety of methods in order to estimate the mass
of M31. Some of the more recent such estimates are displayed in Table \ref{tab:masstable}. For
summaries of older mass estimates we refer the reader to \citet{Federici90,Federici93,Evans00}.

\begin{table*}
 \caption{Estimates of the total mass of M31 found in the recent literature.}
 \label{tab:masstable}
 \begin{tabular}{llll}
        \hline
        \hline
        Reference               & Mass [$10^{12} M_{\odot}$]      & $R_{\rm max}$                     & Method        \\
        \hline
        \citet{Fardal13}        & $1.9^{+0.5}_{-0.4}$                 & 200 times the critical density& Inferred from the Giant Stream                \\
        \citet{Veljanoski13a}   & $1.2-1.5$                     & within 200 kpc                & Dynamical tracers - 50 outer halo GCs         \\
        \citet{Vandermarel12} & $1.5 \pm 0.4$              & within the virial radius & Timing argument + literature + M33     \\
        \citet{Tollerud12}        & $1.2^{+0.9}_{-0.7}$      & within the virial radius  &  Dynamical tracers - 19 dwarf galaxies \\
        \citet{Watkins10}       & $1.4 \pm 0.4$                 & within 300 kpc                & Dynamical tracers - 23 dwarf galaxies         \\
        \citet{Lee08}           & $1.9-2.4$                     & within 100 kpc                & Dynamical tracers - 504 inner regions GCs     \\
        \citet{Evans03}         & $\sim 1.2$                    & within 100 kpc                & Dynamical tracers - 89 inner regions GCs      \\
        \hline
 \end{tabular}
\end{table*}

The M31 outer halo GCs can serve as kinematic mass tracers and therefore be used to provide an estimate
of the total mass of their host galaxy. One way to do this is to solve the Jeans equations
\citep{BT87}. Since we have found the M31 outer halo GC system to exhibit a significant degree of
rotation, in is necessary to separate the solution into two parts: a rotating and a non-rotating,
pressure supported component, the sum of which comprises the total mass of M31.

Since we have assumed the rotation of the halo GCs to occur only on simple circular orbits, the
rotating mass component, $M_{\rm rot}$, is simply determined via:
\begin{equation}
M_{\rm rot} = \frac{R_{\rm max}V^{2}_{\rm max}}{G}
\label{eq:massrot}
\end{equation}
where $V_{\rm max} \equiv A$ is the rotation amplitude of the GC system, $R_{\rm max}$ is the
projected radius of the outermost GCs in the considered sample, and $G$ is the gravitational
constant.

In order to determine the pressure supported mass contribution $M_{\rm pr}$, we use the solution to
the non-rotating Jeans equations proposed by \citet{Evans03}, dubbed the Tracer Mass Estimator (TME):
\begin{equation}
M_{\rm pr} = \frac{C}{GN}\sum\limits_{i=1}^{N}\left(v_i-v_{\rm sys}\right)^{2}R_i
\label{eq:tme}
\end{equation}
where $R$ is the projected radius from the M31 centre for a given GC, $v$ is the radial velocity of
that GC with the rotational component removed, and $N$ is the total number of clusters in the sample
under consideration. The index $i$ loops over each GC in the sample. The constant $C$ is dependent on the
shape of the underlying gravitational potential, the radial distribution of the tracers and the
anisotropy of the system. Here we assume that the M31 outer halo system is spherical and isotropic,
and therefore $C$ takes the form of:
\begin{equation}
C = {4(\alpha\!+\!\gamma) \over \pi}
   {4\!-\!\alpha\!-\!\gamma\over 3\!-\!\gamma}
   {1\!-\!(r_{\rm in}/r_{\rm out})^{3-\gamma} \over 1\!-\!(r_{\rm in}/r_{\rm out})^{4\!-\!\alpha\!-\!\gamma}}.
\label{eq:C}
\end{equation}
In the above definition of C, $r_{\rm in}$ and $r_{\rm out}$ are, respectively, the smallest and largest
deprojected radii of the GCs in the ensemble being studied. For our present sample, the value of
$r_{\rm in}$ is taken to be the distance at which the halo begins to dominate, i.e. 30 kpc. The value
of $r_{\rm out}$ is chosen to be 200 kpc to reflect the measured radius of MGC1, which is the most
remote known M31 cluster \citep{Mackey10a}.

As required by the TME, the GC radial volume density distribution is approximated by a power-law,
the index of which in the case of spherical symmetry is found to be $\gamma\approx 3.34$
(Mackey et al. 2014, in prep.). Hence, in the case of an isothermal halo potential, for which the
$\alpha$ parameter in Equation \ref{eq:C} is zero, we find the total mass enclosed within
200 kpc from the centre of M31 to be $1.6\pm0.2 \times 10^{12}M_{\odot}$. If we assume a NFW
profile for the M31 halo \citep{NFW96}, for which $\alpha\approx0.55$, the estimated mass is somewhat
smaller, at $1.2\pm0.2 \times 10^{12}M_{\odot}$.

In our earlier publication \citep{Veljanoski13a}, we used the same method with a sample of 50 GCs
spanning 3D radii between 30 and 200 kpc in order to estimate the total mass of M31.
Here we update this estimate by by enlarging our halo GC sample to contain a total of 72 GCs covering
the same radial range. Our updated estimate of the dynamical mass of M31 is, perhaps not surprisingly,
consistent with our previous value.  Although these are also comparable to the majority of
the dynamical mass estimates found in the literature that sample a similar spatial range (Table \ref{tab:masstable}),
it is important to state that there are a number of possible caveats related to using the TME in the
present situation.

One important issue is that many of the M31 halo GCs appear to be spatially
associated with cold stellar debris features, and indeed in this paper we have demonstrated that
various groups of such GCs have correlated velocities as described in Section \ref{sec:str}.
This has two consequences. First, the TME assumes that the tracer population is in a steady state
equilibrium, which is not necessarily true as many of the M31 halo GCs are likely to be relatively
new arrivals. Second, because groups of GCs have correlated velocities, it is almost certainly not
appropriate to treat all 72 objects that we used for the mass estimate as independent data points.
In this case we are effectively weighting some data points more heavily than others, introducing a
bias which is not understood for this specific case. That said, studies that have explored the presence of
substructure in tracer populations found their results to be biased only by $\sim20\%$ \citep{Yencho06,Deason12}.

We have also assumed, due to a lack of information to the contrary, that the velocity anisotropy of
the GC orbits is zero, while in reality this is unlikely to be true for the whole halo population.
Nonetheless, \citet{DiCintio12} found the anisotropy parameter to have a negligible effect on mass
estimates derived from tracers for which only radial velocity information is available. Finally, there is
no theoretical motivation to assume that the entire dark matter halo of a massive galaxy follows a
single power law, and thus fixing $\alpha$ to a single number might introduce additional biases.
These caveats give rise to systematic uncertainties in our M31 mass estimate that are not
incorporated in the quoted errors, which only contain the statistical uncertainty. Given the
complex nature of the M31 halo GC population, a more reliable mass estimation may well require
a method specifically tailored to this system.

\section{Summary}
In this contribution we presented radial velocity measurements for 78 GCs around M31, 63 of
which have no previous spectroscopic information. Our sample extends from $\sim20$ kpc out to
$\sim140$ kpc in projection, and at least up to 200 kpc in 3D, which enables us to explore the
kinematic properties of the GCs located in the true outer halo of M31.

Our global kinematic analysis detected a significant degree of net rotation
exhibited by the outer halo GC population of M31. Interestingly, this population shares the same
rotation axis and direction as the GCs located in the inner regions of M31, as well as the M31 disk.
We also find evidence for decreasing velocity dispersion as a function of projected distance from
the centre of M31. The dispersion profile for the halo GC population is similar to that of the stellar
halo, consistent with previous observations that the GCs and stars share similar spatial density profiles.

Our measurements further revealed a variety of velocity correlations for the various groups of GCs
that lie projected on top of distinct stellar debris features in the field halo. In particular, several such GC
groups appear to be kinematically cold, possessing velocity dispersions consistent with zero. Simple
Monte Carlo experiments showed that it is highly unlikely that these velocity correlations are due to chance
arrangements, but rather are most likely due to a common origin for the GCs and the stellar substructures.
This further supports the idea that a significant fraction of the M31 halo GC system has an external
origin \citep[e.g.,][]{Mackey10b}. Definitive proof of the association between GCs and their underlying
streams will require matching velocities between the GCs and the stellar members of the underlying
substructure to be robustly established in each case. We highlighted the few cases where such a
correlation can already tentatively be shown to exist.

In light of the clear association between many groups of halo GCs and underlying field substructures,
our finding that the GC population as a whole possesses a substantial degree of coherent rotation out
to very large radii is quite puzzling. It is difficult to reconcile this property with the chaotic accretion of
parent dwarf galaxies into the halo as implied by our kinematic observations and hierarchical galaxy
formation models. We speculate that the solution to this problem may be related to the recent
discovery that many dwarf galaxies, both in the Milky Way and M31, appear to lie in thin rotating
planar configurations such that their angular momenta are correlated. Alternatively, it may be that
most of the outer halo GCs in M31 arrived with one or two relatively large host galaxies; however,
detailed modelling is required to assess whether this scenario is compatible with the observed narrow
stellar debris streams and the main features of the M31 disk.

Finally, using the halo GCs as kinematic tracers, we estimated the total mass of M31 enclosed
within a deprojected radius of 200 kpc via the Tracer Mass Estimator. Even though our value of
$(1.2-1.6)\pm0.2 \times 10^{12} M_{\odot}$ is in agreement with other recent mass estimates which
employed kinematic tracers extending to similar radii as our sample, it is likely to be subject to
several poorly-understood biases due to the various assumptions that are built into the TME.

\section*{Acknowledgments}
We would like to thank the referee, Flavio Fusi Pecci, for a detailed and constructive
report that helped improve the paper. We also thank Luca Ciotti for carefully reading the
manuscript.

ADM is grateful for support by an Australian Research Fellowship (Discovery Projects Grant DP1093431)
from the Australian Research Council. APH was partially supported by Sonderforschungsbereich SFB
881 ``The Milky Way System" of the German Research Foundation.

The WHT is operated on the island of
La Palma by the Isaac Newton Group in the Spanish Observatorio del Roque de los Muchachos of the
Instituto de Astrof\'{i}sica de Canarias.

This paper is based in part on observations obtained at the Gemini Observatory, which is operated
by the Association of Universities for Research in Astronomy, Inc., under a cooperative agreement
with the NSF on behalf of the Gemini partnership: the National Science Foundation
(United States), the National Research Council (Canada), CONICYT (Chile), the Australian
Research Council (Australia), Minist\'{e}rio da Ci\^{e}ncia, Tecnologia e Inova\c{c}\~{a}o
(Brazil) and Ministerio de Ciencia, Tecnolog\'{i}a e Innovaci\'{o}n Productiva (Argentina).
These observations were obtained under programmes GN-2010B-Q-19,
GN-2011B-Q-61, GN-2012B-Q-77, and GN-2013B-Q-66.

\bibliographystyle{mn2e}
\bibliography{masterRef.bib}

\bsp

\label{lastpage}

\end{document}